\documentclass[jgrga,arxiv,times]{agutex4}
\usepackage{url} 
\usepackage{lineno}
\usepackage{graphicx}
\usepackage{mathptmx}
\usepackage{placeins, mathtools, caption}

\setkeys{Gin}{draft=false}

\authorrunninghead{ABEN ET AL.}
\titlerunninghead{OFF FAULT DAMAGE}

\authoraddr{F. M. Aben, Department of Earth Sciences, University College London, Gower Street, London WC1E 6BS, UK. (f.aben@ucl.ac.uk)}

\begin{document}

\title{Off-fault damage characterisation during and after experimental quasi-static and dynamic rupture in crustal rock from laboratory $P$-wave tomography and microstructures.}

\authors{Franciscus M. Aben,\altaffilmark{1}
Nicolas Brantut\altaffilmark{1}, and
Thomas M. Mitchell\altaffilmark{1}}
\altaffiltext{1}{Department of Earth Sciences, University College London, London, UK}


\begin{abstract}
Elastic strain energy released during shear failure in rock is partially spent as fracture energy $\Gamma$ to propagate the rupture further. $\Gamma$ is dissipated within the rupture tip process zone, and includes energy dissipated as off-fault damage, $\Gamma_\mathrm{off}$. Quantifying off-fault damage formed during rupture is crucial to understand its effect on rupture dynamics and slip-weakening processes behind the rupture tip, and its contribution to seismic radiation. Here, we quantify $\Gamma_\mathrm{off}$ and associated change in off-fault mechanical properties during and after quasi-static and dynamic rupture. We do so by performing dynamic and quasi-static shear failure experiments on intact Lanh\'elin granite under triaxial conditions. We quantify the change in elastic moduli around the fault from time-resolved 3D $P$-wave velocity tomography obtained during and after failure. We measure the off-fault microfracture damage after failure. From the tomography, we observe a localised maximum 25\% drop in $P$-wave velocity around the shear failure interface for both quasi-static and dynamic failure. Microfracture density data reveals a damage zone width of around 10~mm after quasi-static failure, and 20~mm after dynamic failure. Microfracture densities obtained from $P$-wave velocity tomography models using an effective medium approach are in good agreement with the measured off-fault microfracture damage. $\Gamma_\mathrm{off}$ obtained from off-fault microfracture measurements is around 3~kJm$^{2}$ for quasi-static rupture, and 5.5~kJm$^{2}$ for dynamic rupture. We argue that rupture velocity determines damage zone width for slip up to a few mm, and that shear fracture energy $\Gamma$ increases with increasing rupture velocity.
\end{abstract}

\begin{article}

\section{Introduction}
During shear failure in rock, stored elastic strain energy is partly released as radiated energy $E_\mathrm{r}$ (i.e., seismic waves) and mostly dissipated on and around the fault interface as latent heat and new fracture surface area through a plethora of dissipative processes. Dissipated energy is typically partitioned into frictional work and breakdown work, where frictional work $E_\mathrm{f}$ is the work done to overcome the residual friction on the fault interface during sliding. Breakdown work $W_\mathrm{b}$ is a collective term of energies dissipated in addition to $E_\mathrm{f}$, and primarily includes dissipative processes that reduce the strength of the fault interface towards the residual friction. This includes comminution, flash heating \citep{brantut17b}, and thermal pressurisation \citep{viesca15}, but also includes energy dissipated towards propagating the rupture tip, and energy dissipated by deformation outside the principal slip zone (off-fault deformation). For earthquakes, $E_\mathrm{r}$ and $W_\mathrm{b}$ can be determined from seismological data \citep{tinti05, kanamori06}, where $W_\mathrm{b}$ varies from $10^{2}$ to $10^{8}$~Jm$^{-2}$ as a function of total coseismic slip \citep{abercrombie05, viesca15}. As the strength evolution of the fault during failure cannot be determined directly from seismological data, a slip-weakening law is typically assumed to determine a slip-weakening distance $\delta_{0}$, at which the fault has reached its residual frictional strength. Seismological estimates for $W_\mathrm{b}$ do not discriminate between energy dissipated to propagate the rupture, $\Gamma$, and the remaining breakdown work ($W_\mathrm{b} - \Gamma$). $\Gamma$ is called the shear fracture energy and is the energy dissipated within a process zone surrounding the rupture tip to overcome cohesion of the material and propagate the rupture by a unit area \citep{freund90}. $\Gamma$ is dissipated in a volume around the rupture tip, and may therefore include an off-fault component $\Gamma_\mathrm{off}$ in addition to the component of $\Gamma$ dissipated to form the fault interface or principal slip zone. Measurements for material parameter $\Gamma$ are of the order of $10^{4}$~Jm$^{-2}$ for initially intact crystalline low porosity rock under upper crustal conditions \citep{wong82,wong86,lockner91,aben19}, which may be considered an upper bound for pre-existing fault zones often comprised of damaged and altered rock. As $\Gamma$ is dissipated earliest during shear failure \citep{barras20}, its constituent dissipative processes may affect slip weakening processes in the wake of the rupture tip process zone -- and may affect the remainder of $W_\mathrm{b}$. We here aim to quantify the off-fault component of the fracture energy, $\Gamma_\mathrm{off}$. 
 
Off-fault deformation during shear failure, mainly fracturing and subsidiary slip, is created by transient off-fault stresses near the rupture tip \citep{andrews76a,poliakov02,rice05} and by increasingly larger off-fault stresses arising from progressive slip along rough faults \citep{chester00, dieterich09}. Energy dissipated by off-fault deformation in the rupture tip process zone $\Gamma_\mathrm{off}$ is one of $\Gamma$'s constituent energy sinks. During shear failure, off-fault deformation caused directly by the stress concentration around the rupture tip as part of $\Gamma_\mathrm{off}$ precedes most of the off-fault deformation from slip on a rough fault, since the amount of slip within the rupture tip process zone is negligible. Off-fault deformation, and particularly off-fault fracturing, changes the mechanical and hydraulic properties of fault damage zone rock, and thus the constituent dissipative processes of $\Gamma_\mathrm{off}$ affect fault damage zone properties at an early stage during shear failure \citep{aben20b}. This can have a feedback on rupture, slip, and ground motion; rupture simulations show that reduced mechanical properties in the fault damage zone affect fault slip \citep{cappa14} and slip velocity \citep{andrews76a, andrews05, dunham11a}. Due to fracturing near the rupture tip the pore volume increases and causes, under partially undrained conditions, a local pore fluid pressure drop and an increase in effective pressure on the fault \citep{brantut20}. This can stabilise dynamic rupture \citep{martin80} and slip \citep{segall95,segall10}. Changes in hydraulic properties from off-fault fracture damage close to the fault interface have an effect on slip-weakening mechanisms that act in the wake of the rupture tip, such as thermal pressurisation \citep{brantut18c}. The dynamic reduction of elastic moduli in the fault damage zone causes high frequency content in the radiated ground motion \citep{thomas17}, and can be a substantial additional source of seismic radiation \citep{ben-zion09}. It is therefore crucial to 1): Quantify $\Gamma_\mathrm{off}$, and 2): Quantify the changes it imposes on off-fault mechanical properties.

A measurement of total off-fault fracture surface area created in the rupture tip process zone gives an estimate for the cumulative fracture surface energy necessary to create them. This gives a lower bound for $\Gamma_\mathrm{off}$, as energy dissipated as latent heat during off-fault fracturing (i.e., slip on the fractures) remains unknown. Along strike-slip faults, this approach has yielded estimates for the total off-fault dissipated energy \citep{chester05, rockwell09}. However, fractures observed in exhumed fault damage zones originate from either rupture tip stress concentrations, stresses generated by slip on a rough fault during shear failure, or quasi-static stresses \citep{mitchell09}, and were healed and overprinted by numerous shear failure events. This complicates quantification of $\Gamma_\mathrm{off}$ from the geological record. Off-fault fracture damage induced by shear failure under controlled conditions in the laboratory circumvents some of these complications, allowing for a microstructural description \citep{wawersik71, reches94} and quantification of fracture damage zones \citep{moore95, zang00} associated to a single failure event. \citet{moore95} estimated the cumulated surface energy in the fracture damage zone around a `frozen' quasi-static rupture front in granite, where slip on the fault was negligible, yielding a lower bound for $\Gamma_\mathrm{off}$. A dynamically propagating rupture tip is expected to create a larger area of fracture damage, as the stress field around a propagating rupture tip is distorted with increasing rupture velocity \citep{poliakov02}, and we therefore expect $\Gamma_\mathrm{off}$ to increase as well. 

$\Gamma_\mathrm{off}$ can also be obtained from the change in stored elastic strain energy in the rupture tip process zone, with the underlying assumption that the change in elastic compliance is caused by off-fault fracturing. A reduction in elastic compliance is measured as a drop in seismic wave speeds, making them an attractive and cost-efficient proxy for large scale monitoring of fracture damage structures in fault zones \citep{mooney86, rempe13, hillers16, qiu17}. To date, high resolution geophysical measurements of wave speeds from dense arrays \citep{ben-zion15} have given static snapshots of the fault damage zone structure, but not the coseismic velocity drop necessary to obtain the total coseismic off-fault dissipated energy, let alone $\Gamma_\mathrm{off}$. Laboratory-scale seismic tomography of the $P$-wave velocity structure \citep{brantut18} obtained from ultrasonic data measured during quasi-static shear failure experiments does give the change in effective elastic moduli during rupture needed to calculate $\Gamma_\mathrm{off}$ \citep{aben19}, yielding a similar value for $\Gamma_\mathrm{off}$ to that calculated from fracture surface area by \citet{moore95}. There are, to our knowledge, no measurements of $\Gamma_\mathrm{off}$ for dynamic shear ruptures yet, either from microstructures or from a change in elastic moduli.

The changes in off-fault mechanical properties induced by shear rupture cannot be assessed directly from the scalar quantity $\Gamma_\mathrm{off}$, but the two approaches outlined above to estimate $\Gamma_\mathrm{off}$ also provide the changes in elastic moduli and the microfracture density. These two physical properties can be reconciled using effective-medium theory models for cracked solids \citep[e.g.,][]{gueguen11}, which are an important tool for obtaining information on physical and hydraulic properties such as fracture density \citep{sayers95}, porosity, and permeability \citep{gavrilenko89}. These physical parameters are key in studying the feedback between rupture and slip. Effective-medium approaches have been tested in the laboratory on deformed samples, where effective elastic moduli were measured by active ultrasonic surveys \citep[e.g.,][]{schubnel03}. The path-averaged wave velocities obtained from these surveys are representative for fracture damage only when fractures are homogeneously spread throughout the sample. In laboratory shear failure experiments, fracture damage is localised around the fault interface and so path-averaged velocities cannot be used. Instead, recent advances in syn-deformation laboratory tomography techniques \citep{brantut18, stanchits03} can be employed for the use of effective-medium models, so that changes in physical properties can be quantified in situ. 

Here, we assess $\Gamma_\mathrm{off}$ for dynamic and quasi-static rupture in granite following the two approaches outlined above. To do so, we perform three types of shear failure experiments in the laboratory: Shear failure by quasi-static rupture, by dynamic rupture, and by partly quasi-static and partly dynamic rupture (from here on referred to as `mixed rupture'). We quantify the change in mechanical properties around the fault caused by shear failure from time-resolved 3D $P$-wave velocity tomography models. These were obtained during and after quasi-static rupture and after dynamic rupture and mixed rupture. We also quantify the off-fault microfracture damage after dynamic and quasi-static shear failure from microstructural observations. An effective-medium approach is used to obtain microfracture densities from the 3D $P$-wave velocity models, which are compared with the measured microfracture densities. We then determine a damage zone width for the quasi-statically and dynamically failed samples. These estimates for damage zone width are compared to the expected damage zone width from the stress field around a propagating rupture tip \citep{poliakov02} and from the off-fault stresses induced by slip along a rough fault \citep{chester00}. We then obtain $\Gamma_\mathrm{off}$ from measuring the cumulative off-fault fracture surface energy within the damage zone. These measurements are complementary to $\Gamma_\mathrm{off}$ derived from changes in effective elastic moduli by \citet{aben19} for a quasi-static rupture. Last, we discuss the implications of our results to the energetics of earthquake rupture. 

\begin{table*}
\centering
{
\begin{tabular}{l | c c c c c}
sample number								& LG1 & LN4 & LN5$^\dagger$ & LN7 & LN8 \\ \hline
type of experiment 							& dynamic$^{\ddagger}$ & mixed & quasi-static & dynamic & mixed$^{\ddagger}$ \\
nr. of time intervals							& $8$ & - & $38$ & - & 22 \\
number of AE events							& $2215$ & - & $11134$ & - & $9844$ \\ \hline
survey arrival time [$\mu$s] 					& $1$ & - & $1$ & - & $1$\\ 
anisotropy parameter [-] 						& $0.01$ & - & $0.01$ & - & $0.01$\\
\textit{a priori} velocity model [$\log{\textrm{(m/s)}}$] 	& $0.02$ & - & $0.02$ & - & $0.01$\\
AE arrival time [$\mu$s] 						& $2$ & - & $2$ & - & $2$\\
AE source location [mm] 						& $2$ & - & $2$ & - & $2$\\
AE origin time [$\mu$s] 						& $2$ & - & $2$ & - & $2$\\
correlation length [mm]						& $25$ & - & $25$ & - & $25$ \\  \hline
microstructural analysis						& - & - & 123 images & 134 images & -\\ \hline
slip $\delta$ [mm]							& 2.88 & 1.93 & 0.83 & 3.22 & 2.44 \\
\end{tabular}}
\caption{Sample table with experiment type, number of time intervals, number of AE events used for tomographic inversion, tomographic inversion parameters (covariances and correlation length), and number of SEM images used for microstructural analysis. $^\dagger$From \citet{aben19}. $^{\ddagger}$Accumulated slip by reloading after failure.}
\label{tab:1}
\end{table*}

\section{Materials and methods}
\subsection{Experiments}
Three different types of failure experiments were performed on intact 100~mm by 40~mm diameter Lanh\'elin granite cylinders (from Brittany, France) at 100~MPa confining pressure (Table \ref{tab:1}): Failure by dynamic rupture, failure by quasi-static rupture, and failure by part quasi-static rupture and part dynamic rupture named mixed rupture. The experiments were performed at nominally dry conditions in a conventional oil-medium triaxial loading apparatus at University College London \citep{eccles05}. Axial load was measured by an external load cell corrected for friction at the piston seal. Axial shortening was measured by a pair of Linear Variable Differential Transducers (LVDTs) outside the confining pressure vessel, corrected for the elastic shortening of the piston. 

The samples were equipped with two pairs of axial-radial strain gauges. The samples were placed in a rubber jacket equipped with 16 piezoelectric $P$-wave ($V_\textrm{P}$) transducers. Ultrasonic signals were amplified to 40~dB before being recorded by a digital oscilloscope (50~MHz sampling frequency). All signals consisted of 4096 data points, equivalent to an 82~$\mu$s time interval. Active ultrasonic velocity surveys were performed every 5 minutes, where all 16 piezoelectric transducers were sequentially used as a source, while the other transducers recorded the resulting waveforms. 1~MHz pulses were produced by exciting the source transducer with a 250~V signal. The signal-to-noise ratio was improved by stacking the recorded waveforms from six of these pulses per transducer. Between surveys, acoustic emissions (AE) were recorded on 16 channels, provided that the AE signal amplitude was above 250~mV on at least two channels within a 50~$\mu$s time interval. The digital oscilloscope stored up to four sets of AE waveforms per second. 

Dynamic rupture was achieved by setting a constant shortening rate equivalent to an axial strain rate of $10^{-5} \textrm{ s}^{-1}$ until dynamic shear failure. Quasi-static rupture was achieved by suppressing dynamic rupture via monitoring the AE rate, following the approach of \citet{lockner91}. When the acoustic emission rate showed a marked increase -- a precursor to dynamic rupture -- the axial load on the sample was decreased by reversing the displacement direction of the piston. For mixed rupture experiments, the rupture was controlled for about half the stress drop between the sample's peak stress and its residual frictional strength. The rupture was allowed to propagate dynamically for the remainder of the stress drop. After failure, one sample failed by dynamic rupture and one sample failed by mixed rupture were reloaded up to their residual frictional strength (Table \ref{tab:1}), which resulted in some additional stable sliding along the fault. 

Poisson's ratio of the intact rock $\nu_{0}$ was determined from the ratio of the axial and radial strain during axial loading in the elastic regime. The intact Young's modulus $E$ was derived from the differential stress versus axial displacement curves measured during axial loading in the elastic regime. 

\subsection{Analysis of ultrasonic data and $P$-wave tomography}
The \textit{FaATSO} code by \citet{brantut18} was used for tomographic inversion of the active ultrasonic surveys and AE arrival times. Prior to tomographic inversion, the ultrasonic waveforms recorded during the experiments were processed. Time of flight for all sensor combinations were picked for the first active ultrasonic survey of the experiment, and arrival times for subsequent surveys were extracted using an automated cross-correlation technique \citep[e.g.,][]{brantut14} with a precision of about 0.05~$\mu$s. From these, path-averaged velocities were calculated between sensor pairs. These ray paths are oriented at 90$^{\circ}$ (i.e., horizontal), 58$^{\circ}$, 39$^{\circ}$, and 28$^{\circ}$ angles to the loading axis of the sample. AE arrival times and source locations were obtained in three steps: 1) The first arrivals of the AE waveforms were automatically picked, and AE source locations were calculated using their arrivals in conjunction with a transverse isotropic velocity model based on the most recent ultrasonic survey. 2) The AE events were subjected to a quality test, where AEs with a source location error above 5~mm were discarded. 3) The automatically picked arrival times of the remaining AEs were subjected to an interactive visual check -- arrival times were improved or removed when the difference between the automatically picked arrival time and the theoretical arrival time for the calculated source location was too large. 4) The AE source locations were recalculated based on the inspected arrival time dataset and the same source location error criterium was applied. 

The \textit{FaATSO} code treats the arrival times of the ultrasonic surveys and the AE arrival times as the observed data. The model parameters are the AE source locations and origin times, and the horizontal $P$-wave velocity and anisotropy in voxels of $5\times5\times5$~mm that cover the sample volume. The algorithm allows for vertical transverse isotropy for each voxel (i.e., the vertical velocity is independent from the horizontal velocity). $V_\textrm{P}$ anisotropy is expressed as the ratio $(V_{\textrm{P}}^{\textrm{v}} - V_{\textrm{P}}^{\textrm{h}}) / V_{\textrm{P}}^{\textrm{h}}$, where $V_{\textrm{P}}^{\textrm{h}}$ and $V_{\textrm{P}}^{\textrm{v}}$ are the horizontal and vertical $P$-wave velocities, respectively. To make predictions of the observed data based on the model parameters, a 3D anisotropic ray tracer is used (i.e., Eikonal solver) \citep{brantut18}. The inverse problem is solved using a quasi-Newton inversion algorithm \citep{tarantola05}, and is constrained by a set of standard deviations that describe Gaussian variances on the observed data (Table \ref{tab:1}). The variance on the model parameters (AE source locations, velocity, and anisotropy) are also Gaussian, expressed by standard deviations (Table \ref{tab:1}). For the velocity and anisotropy, there is a covariance between voxels that is a function of the variance and a correlation length \citep{brantut18}. Through the covariance for velocity and anisotropy, the correlation length smooths heterogeneities in the inversion results. 

The observed data was divided in a number of time intervals (Table \ref{tab:1}) with varying duration, each containing roughly 300 AEs, for which we performed the inversion. The AE source locations were used as an \textit{a priori} model parameter. For the remaining \textit{a priori} model parameters, $V_{\textrm{P}}^{\textrm{h}}$ and anisotropy, we used two structures: 1) A homogenous vertical transverse isotropic (VTI) \textit{a priori} velocity structure derived from the most recent ultrasonic survey in each time interval, and 2): An inherited \textit{a priori} velocity structure from the inversion results of the preceding time interval, except for the first time interval where a homogenous VTI \textit{a priori} structure was used. The quality of the inversion results was tested by comparing both sets of inversion results (see Text S1).

A clear tomographic image during dynamic rupture could not be achieved by inversion of pre- and syn-rupture AE events, because the number of recorded syn-rupture events is too low due to the limited recording capacities of the acquisition system, and pre-rupture events occur at a stage where deformation is not yet localised. We therefore use AE events recorded during reloading of a dynamically failed sample and a sample failed by mixed rupture.

\subsection{Microstructural analysis}
\begin{figure*}
\centering
\includegraphics[scale = 0.9]{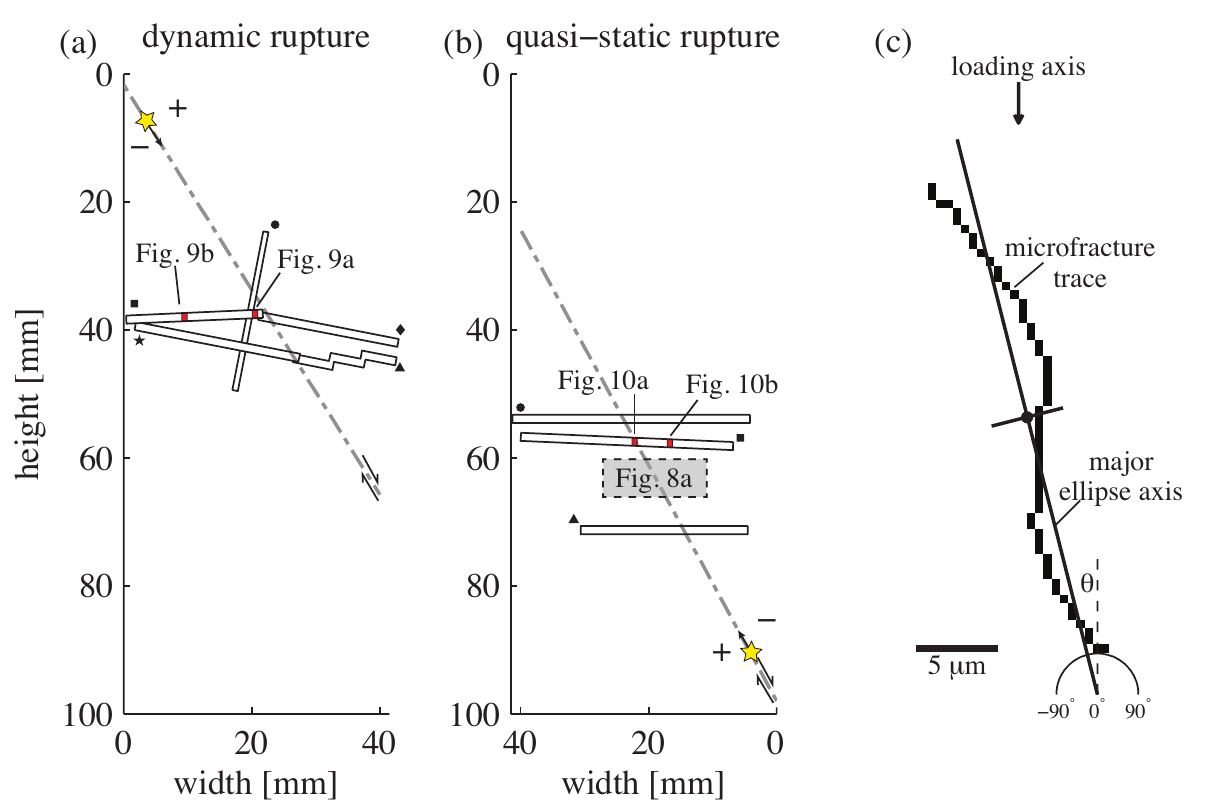}
\caption{Sketch of fault-perpendicular slice through the centre of a sample showing locations of the transects along which SEM images have been obtained (to scale), for \textbf{(a)}: Dynamic rupture, sample LN7, and \textbf{(b)}: Quasi-static rupture, sample LN5. Each transect is assigned a symbol corresponding to the fracture density data in Figures \ref{fig:9}d and \ref{fig:10}d. The trace of the main fault plane is shown by the dashed gray line. The red windows indicate the locations of individual SEM images shown in Figures \ref{fig:9}a, b,  and \ref{fig:10}a, b. The compressive and tensile lobes in the rupture tip process zone are marked by $+$ and $-$ signs, the star and arrow on the fault marks the approximate location of rupture nucleation and propagation direction. Location of the optical microscopy image in Figure \ref{fig:8}a indicated in (b). \textbf{(c)}: Trace of a fracture segment, and the minor and major axis of an ellipse fitted around the segment. The fracture segment length used to calculate $\rho_\mathrm{frac}$ is given by the number of its constituent pixels. The angle $\theta$ between the major ellipse axis and the loading axis gives the fracture orientation. }
\label{fig:1}
\end{figure*}

Polished thin sections oriented perpendicular to the main fault interface were cut from epoxied post-mortem samples that failed by dynamic rupture and by quasi-static rupture. The thin sections were studied by optical microscopy and by scanning electron microscope (SEM), from the latter we obtained back-scatter electron (BSE) grayscale images along three transects through the centre of the sample (Figure \ref{fig:1}a, b; Table \ref{tab:1}). The images were taken at a $100\times$ magnification and cover a $1.0$ mm$^{2}$ area. The pixel dimension is 0.5 by 0.5~$\mu$m. 

We obtained the traces of microfractures as follows: Microfractures are revealed as low grayscale value features in the SEM pictures, because they are empty or filled with low density epoxy. The microfractures may be traced by hand, but given the large number of SEM images, we elected to use a semi-automated image analysis technique. Both methods are prone to user errors, but the errors from semi-automated image analysis are more consistent in all images so that analysis within the dataset itself is more reliable. The microfractures can be isolated by using a grayscale threshold, but this approach will isolate pores in addition to open fractures, and will exclude pixels of low aperture fractures because they partly overlap with higher density wall rock, which increases the absolute grayscale value. Fracture recognition from sharp grayscale contrasts (i.e., edge detection) is more sensitive to low aperture fractures, but will also recognise pores and sharp grain boundaries between different minerals. Here, we isolate microfractures based on fracture aperture, so that larger aperture pores can be excluded. To do so, we use the median filter technique used by \citet{griffiths17}, and incorporate their approach in the newly developed fracture tracing code \textit{Giles} (fracture tracin\underline{\textbf{G}} by med\underline{\textbf{i}}an filter, ske\underline{\textbf{l}}etonisation, and targ\underline{\textbf{e}}ted clo\underline{\textbf{s}}ure, freely available on https:\slash \slash github.com\slash FransMossel\slash  Giles\_fracturetracing.git). The median filter obtains a median grayscale value for a predefined window of pixels around a target pixel, and assigns this median value to the target pixel. The entire image is subjected to this action. If the predefined window is larger than the fracture aperture and smaller than the aperture of pores, it ascribes a median grayscale value to a pixel in the fracture that is much higher than the original value, but pixels that represent pores or grains do not significantly change \citep{griffiths17}. The difference between the original grayscale values and the median filtered values is thus much higher in microfractures than in surrounding grains and pores. The image of this difference is therefore binarised. Small gaps between fracture traces in the binarised image are closed with a dilation-erosion action. The binary image is skeletonised, reducing the width of the trace to a single pixel, followed by targeted closure of gaps between traces that have the same orientation. Small residual branches on the fracture traces are an artefact of the skeletonisation process, and are removed by a pruning algorithm similar to that used by \citet{griffiths17}. A visual check and, when necessary, adjustment of the user-defined parameters, is imperative to ensure reasonable results from the fracture tracing code. See Text S1 for more details on the image analysis steps. 

The end result of image processing using \textit{Giles} is a binary image with microfracture traces of single pixel width. Fractures below 3 to 9~$\mu$m in length (depending on the size of the median filter window) were not traced. Since we are primarily interested in off-fault damage, we manually removed fracture traces in gouge-filled zones and zones of cataclasite. We do not define individual fractures, because this requires manual unravelling of the microfracture network that would give arbitrary results for a well-connected fracture network where a clear fracture hierarchy is missing. Instead, we analyse fracture segments, which are defined as pixels connected to only two neighbours. Fracture segments are separated by fracture intersections, which are pixels with three or more neighbours.

We obtained 2D fracture orientations for each fracture segment by fitting an ellipse around a segment and measuring the angle $\theta$ between the major axis of the ellipse and the sample axis (Figure \ref{fig:1}c). The absolute cumulative fracture length in an image is given by the number of pixels used for the fracture traces. Off-fault fracture density $\rho^{\textrm{frac}}$ (in mm/mm$^{2}$) was obtained for each image by dividing the total fracture length in an image with the surface area of that image. The SEM image transects span both sides of the fault zone, which experienced different transient stresses in the rupture tip process zone. The transient off-fault stresses are tensile on the side of the fault where the direction of slip is opposite to the rupture propagation direction, and compressive on the other side of the fault. Based on the migration of AE source locations over time, which indicates the rupture propagation direction, we identified the tensile and compressive sides of the fault (Figure \ref{fig:1}a, b).

\section{Results}
\subsection{Experiments}
\begin{figure*}
\centering
\includegraphics[scale = 0.9]{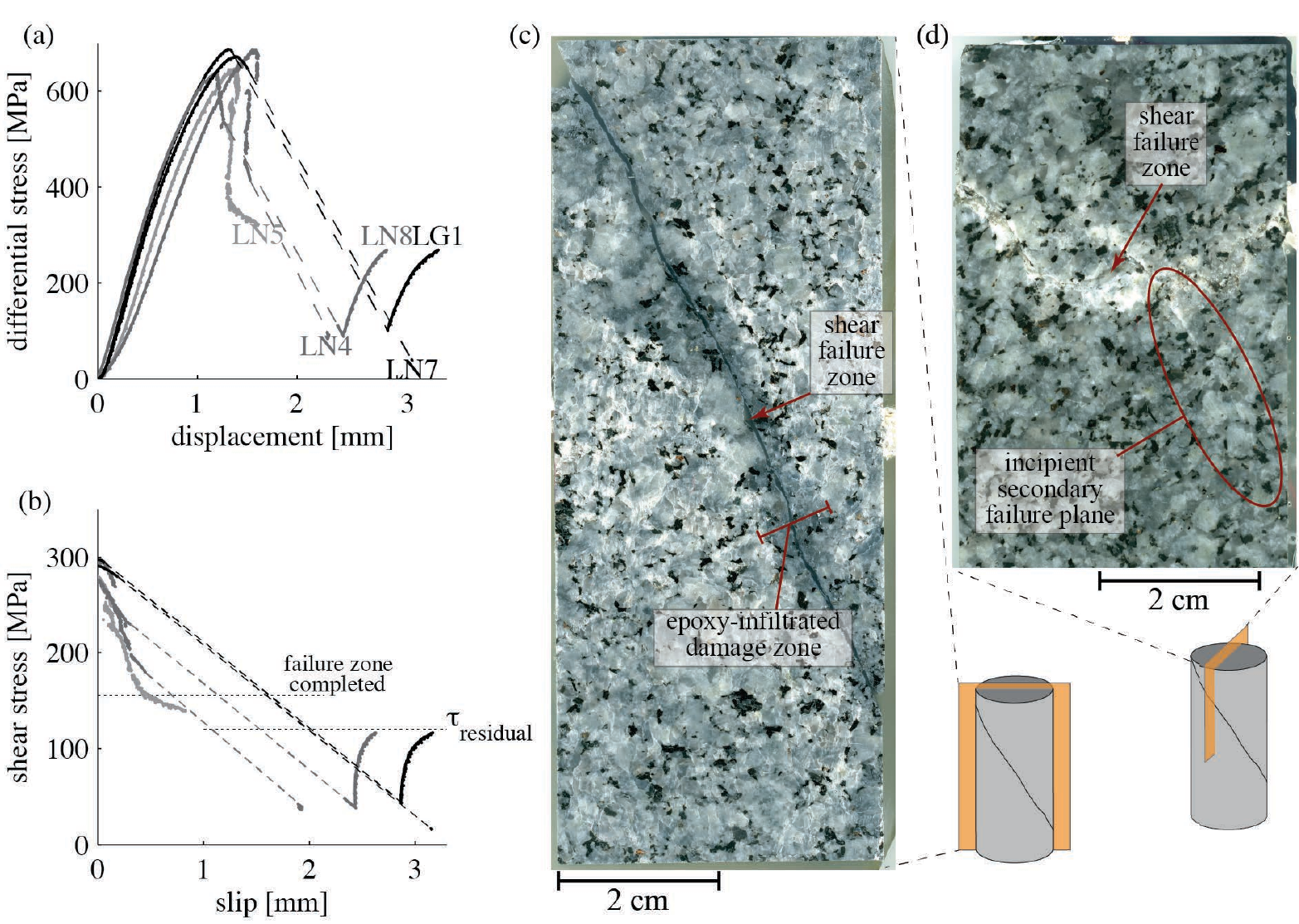}
\caption{\textbf{(a)}: Stress-displacement curves for Lanh\'elin granite samples subjected to dynamic rupture (LN7 and LG1, black), to quasi-static rupture (LN5, light gray), and to mixed rupture (LN4 and LN8, dark gray). The displacement and stress drop caused by dynamic failure are shown as dashed intervals. \textbf{(b)}: Shear stress versus slip curves for all failure experiments. Curves follow the same colour coding as in (a).The shear stress at which the failure zone was completely formed is indicated, and $\tau_\mathrm{residual}$ gives the residual shear stress to which the curves converge. \textbf{(c)}: Polished section of dynamically failed sample LN4, oriented perpendicular to the failure zone. Note that epoxy has penetrated the failure zone and part of the damage zone (darkened area), but has not penetrated the intact rock near the edge of the sample. \textbf{(d)}: Polished section of mixed ruptured sample LN8, oriented parallel to the main failure zone. A perpendicular incipient failure plane is visible. Surface has been epoxied prior to polishing, so that the damage zone is less apparent compared to (c). }
\label{fig:2}
\end{figure*}

The samples reached a peak differential stress of 660 to 700~MPa, followed by the onset of fault localisation and rupture propagation (Figure \ref{fig:2}a). Frictional sliding -- and thus the completion of rupture -- commenced between 360 and 350~MPa, based on the flattening of the stress-displacement curve and the spread of the AE source mechanisms across the entire slip surface during quasi-static rupture. The post-failure residual strength is around 300~MPa, as shown by the converging stress-displacement curves of the quasi-static, dynamic, and mixed experiments, of which the latter two approached the residual frictional strength from a lower differential stress by reloading of the sample. 

Visual inspection of the samples after the deformation experiment revealed a single shear failure zone (Figure \ref{fig:2}c), except for sample LN8 that includes an incipient secondary fault plane without noticeable displacement in addition to the through-going shear failure zone (Figure \ref{fig:2}d). All through-going failure zones are oriented approximately at 30$^{\circ}$ relative to the compression axis. Using this fault angle, we resolved the average shear stress on the fault plane from the differential stress and confining pressure (Figure \ref{fig:2}b). The rupture fully traversed the sample and completed the failure zone at about 155~MPa shear stress (Figure \ref{fig:2}b), as measured from the quasi-static rupture, and the residual frictional strength $\tau_\mathrm{residual}$ is around 120~MPa, as shown by the converging stress-strain curves for quasi-static, dynamic, and mixed ruptures (Figure \ref{fig:2}b). The slip on the fault $\delta$, calculated from the axial displacement data corrected for machine stiffness and for the stiffness of the intact rock, was 0.83~mm at the end of the quasi-static rupture experiment, 2.88 to 3.22~mm  after dynamic failure, and 1.93 to 2.44~mm after dynamic failure in mixed rupture experiments (Table \ref{tab:1}). Additional slip of 0.19~mm and 0.29~mm was accumulated by reloading samples LG1 and LN8, respectively, after dynamic failure. 

A Young's modulus $E = 88$~GPa was measured for the intact rock during axial loading above 100~MPa and below about 400~MPa differential stress, and averaged over all experiments. Averaged over all experiments, a Poisson's ratio $\nu_0 = 0.20$ was estimated for intact rock. 

\subsection{Ultrasonic velocity surveys}
\begin{figure}[b]
\centering
\includegraphics[scale=0.85]{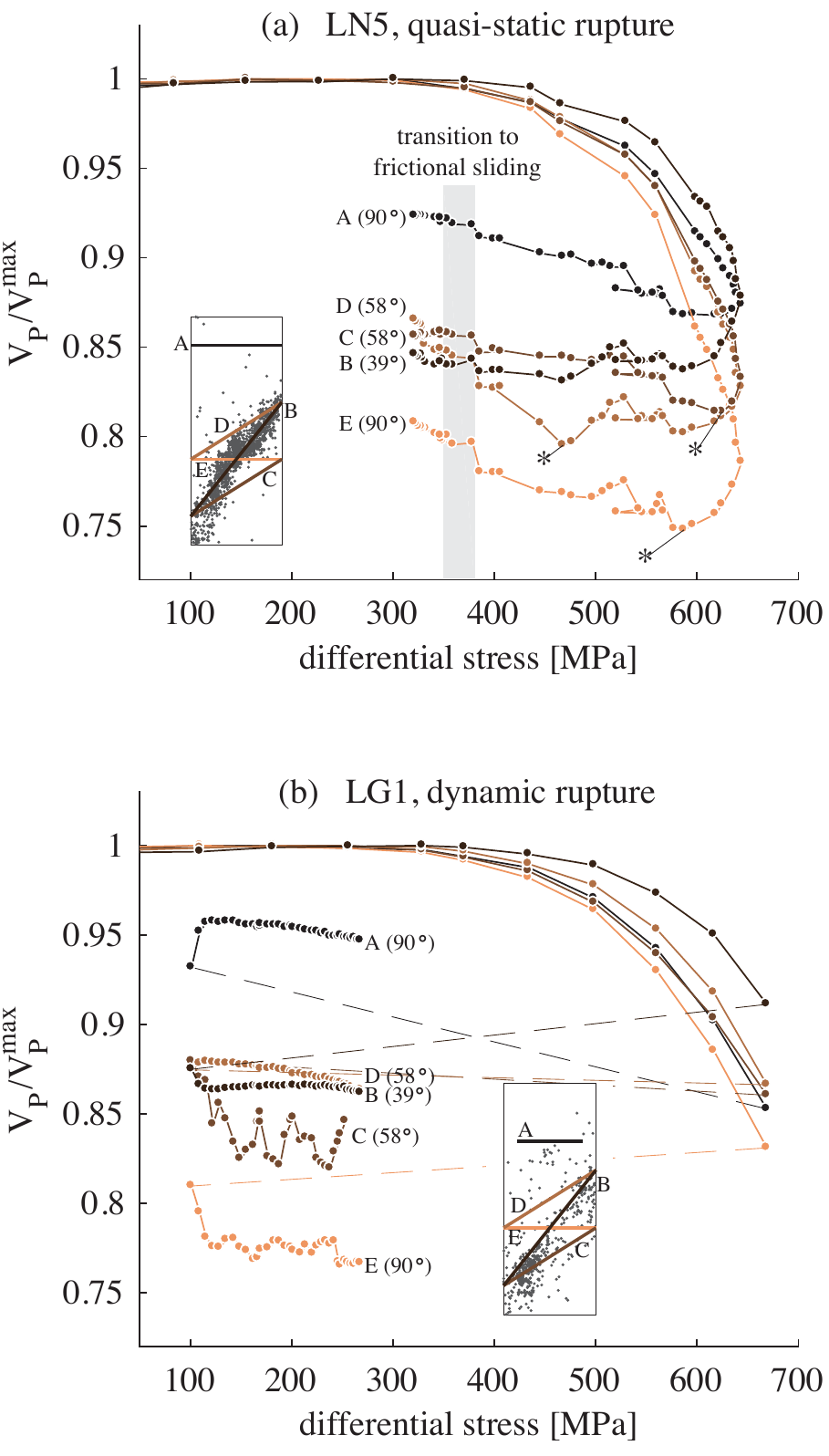}
\caption{\textbf{(a)}: Normalised path-averaged $P$-wave velocity measured during a quasi-static failure experiment (sample LN5) versus differential stress. $P$-wave velocities were obtained from the first $P$-wave arrival of active ultrasonic surveys. The shaded area indicates the transition to frictional sliding, and the asterisks highlight the lowest velocities of three ray paths (see main text). The curves are coloured similar to their locations shown in the cross-section through the centre of the sample (inset) where the fault plane, delineated by AE source locations within 2~mm of the cross-section, intersects the cross-section and ray paths at a 45$^\circ$ angle. \textbf{(b)}: Normalised $P$-wave velocity measured before and after a dynamic failure experiment (sample LG1) versus differential stress. The dynamic stress drop during failure is dashed. The curves are coloured similar to their locations shown in the cross-section through the centre of the sample (inset) where the fault plane, delineated by AE source locations within 3~mm of the cross-section, intersects the cross-section and ray paths at a 40$^\circ$ angle. }
\label{fig:3}
\end{figure}

Path-averaged ultrasonic $P$-wave velocities were routinely calculated from the time of flight between two sensors, assuming a straight ray path (i.e., shortest distance) between the sensors. We present the $P$-wave velocity change during deformation with respect to the initial $P$-wave velocity at hydrostatic conditions along 5 straight ray paths at key orientations with respect to the fault plane during a quasi-static failure experiment (Figure \ref{fig:3}a) and a dynamic failure experiment (Figure \ref{fig:3}b). In both samples, ray path~A is perpendicular to the loading axis and located well outside the eventual failure zone. Ray path~B is oriented at 39$^\circ$ to the loading axis, and nearly its entire length is located within the failure zone. Ray paths~C and~D, both at a 58$^\circ$ to the loading direction, intersect with the two extremities of the fault zone and run sub-parallel to it. Ray path~E, oriented perpendicular to the loading direction, intersects the fault zone in the centre of the sample.

Before the onset of quasi-static rupture, path-averaged $P$-wave velocities along all 5 ray paths increase slightly by 30~m/s up to 6.2~km/s from 0~to~400~MPa differential stress (Figure \ref{fig:3}a), followed by a strong decrease as peak stress is approached. At the peak differential stress, $V_\mathrm{P}$ along ray path~A shows the smallest velocity reduction of about 13\% down to 5.3~km/s. During quasi-static rupture, when differential stress drops from peak stress to about 350~MPa, $V_\mathrm{P}$ along ray path~A recovers by 6\%, and remains stable during sliding between 350-300~MPa differential stress. Ray path~B reveals a drop in $P$-wave velocity of about 13\% at the peak differential stress. Between the peak stress and 600~MPa differential stress, $V_\mathrm{P}$ along ray path~B decreases by an additional 4\%, and remains stable at a total reduction of 17\% for the remainder of the stress drop. After rupture completion at 350~MPa differential stress and the onset of sliding, the $P$-wave velocity along ray path~B recovers by 1-2\%. At the peak stress, $P$-wave velocity along ray path~C and~D dropped by 17\%. $V_\mathrm{P}$ continues to decrease, along~C down to 21\% at 625~MPa differential stress, and along D down to 19\% at 470~MPa (Figure \ref{fig:3}a, asterisks). At the end of the experiment, after frictional sliding, the overall velocity drop along ray paths~C and~D is 14\% and 15\% respectively, which is a velocity recovery of 7\% and 4\% with respect to the minimum observed $V_\mathrm{P}$. The velocity drop along ray path~E was 22\% at the peak stress. Along this ray path, we observe the strongest reduction in $P$-wave velocity of about 26\% down to 4.6~km/s at a differential stress of 590~MPa during quasi-static failure (Figure \ref{fig:3}a, asterisks). As failure progresses and differential stress drops further, the velocity recovers so that a 19\% reduction in $V_\mathrm{P}$ is measured at the end of the experiment. 

Path-averaged $P$-wave velocities before dynamic failure (Figure \ref{fig:3}b) are similar to those before quasi-static rupture. Ultrasonic surveys could not be obtained during dynamic rupture, but were obtained during reloading of the sample after failure. $P$-wave velocity outside the fault zone along ray path~A was reduced by 15\% down to 5.3~km/s prior to dynamic failure from an initial velocity of 6.2~km/s. The dynamic stress drop during failure caused an increase in $V_\mathrm{P}$ of 8 to 11\%, followed by a small decrease during reloading down to 5.9~km/s at 270~MPa differential stress (Figure \ref{fig:3}b). $P$-wave velocity along ray path~B drops from 5.7~km/s (9\% drop) at peak stress to 5.4~km/s (13\% drop) after dynamic failure (Figure \ref{fig:3}b). During reloading, the $P$-wave velocity does not change along wave path~B. Pre-failure $P$-wave velocities along ray paths~C and~D drop by 13--14\% along both wave paths, and recover by 1\% up to 5.4--5.5~km/s after the dynamic stress drop (Figure \ref{fig:3}b). Within the fault zone along ray path~E, the $P$-wave velocity drops during the dynamic stress drop from 5.1~km/s down to 4.8~km/s (22\% reduction, Figure \ref{fig:3}b). $V_\mathrm{P}$ decreases slightly more during reloading of the sample (down to a 23\% reduction). 

Path-averaged $P$-wave velocity changes measured during quasi-static rupture and before and after dynamic rupture are of similar magnitude and show a wide variation in velocity reductions within a single sample, with $V_\mathrm{P}$ reduced by 5 to 24\% at the end of the experiment relative to the intact rock. These variations indicate strong localisation of damage, and the difference between horizontal $V_\mathrm{P}$ (measured perpendicular to the loading axis) and $V_\mathrm{P}$ measured at an angle indicate damage-induced anisotropy. Overall, $P$-wave velocity tends to increase with decreasing differential stress, except for the ray paths located entirely within the fault zone (ray paths~B in Figure \ref{fig:3}a and b). The above analysis assuming straight ray paths reveals very precise changes in path-averaged $V_\mathrm{P}$ thanks to the cross-correlation technique used to extract arrival times. However, changes in path-averaged $V_\mathrm{P}$ do not reveal where along the ray path the $V_\mathrm{P}$ has changed. Therefore, we perform a 3D tomographic inversion, which will lack the precision of the path-averaged velocity changes, but will reveal the location of greatest change in $V_\mathrm{P}$.  

\subsection{$P$-wave tomography}
We first present changes in the horizontal $P$-wave velocity structure introduced by quasi-static, dynamic, and mixed rupture. The results were obtained using an inherited \textit{a priori} velocity model (see section 2.2 and Text S2). We then present the $P$-wave anisotropy inversion results. We detail the difference between the inherited and the homogeneous VTI \textit{a priori} velocity models and the effect of different standard deviations on the model parameters in Supplementary information Text S2. 

\begin{figure*}
\centering
\includegraphics[scale = 0.9]{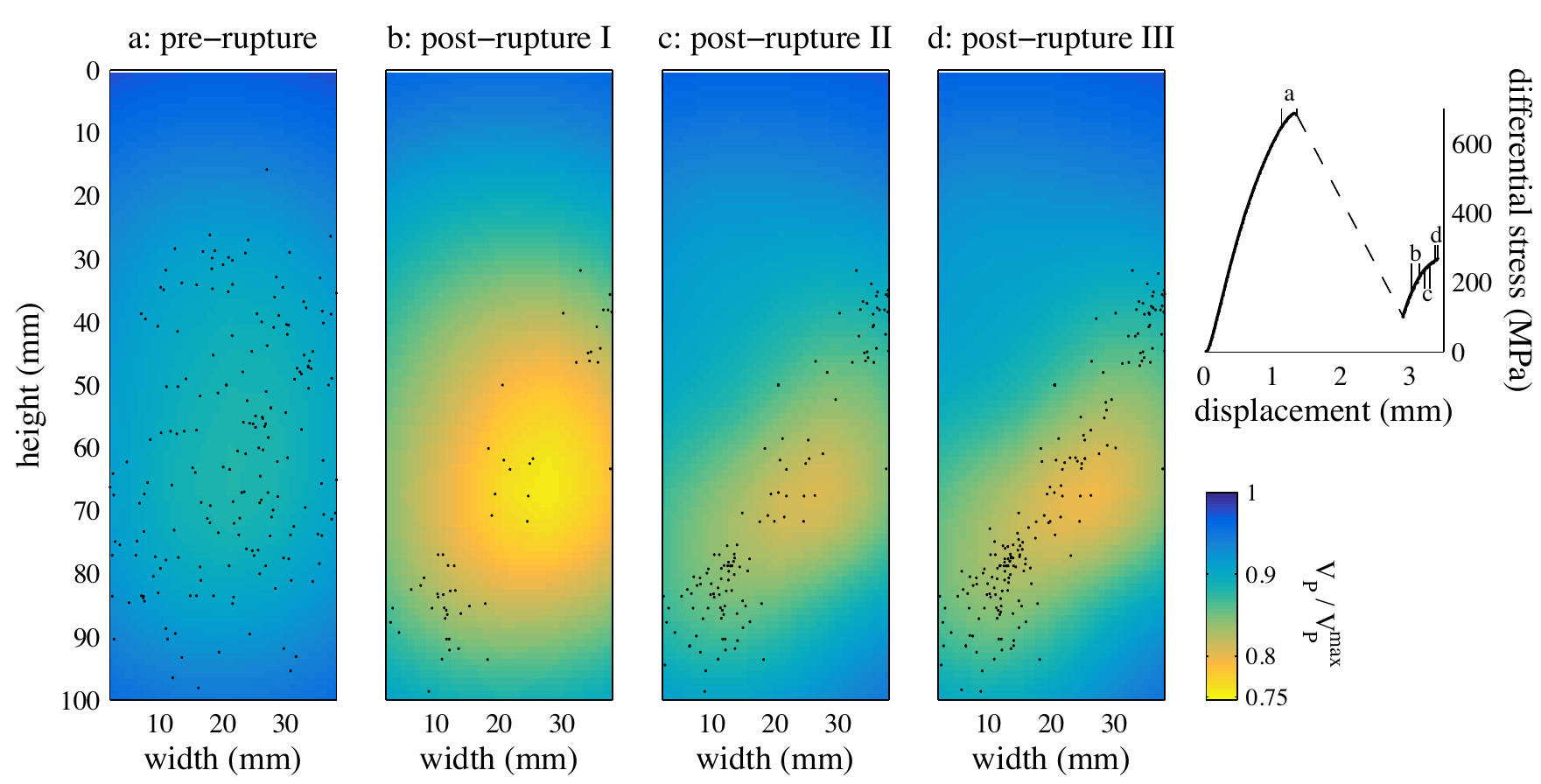}
\caption{Tomographic sections of the horizontal $V_\mathrm{P}$ normalised to the initial velocity through the centre of  sample LG1 (dynamic rupture), perpendicular to the fault. The four slices show time intervals \textbf{(a)} just prior to the localisation of deformation, and \textbf{(b)},\textbf{(c)}, and \textbf{(d)} during three stages of reloading after rupture and slip. The corresponding parts of the stress-displacement curve are indicated on the right. AE source locations up to the time interval shown are projected onto the slice. The AE source locations are within 5~mm distance perpendicular to the slide. In this figure, and Figures \ref{fig:5} and \ref{fig:6}, the seismic velocities are smoothed to a 1~mm resolution and an inherited \textit{a priori} model was used for the inversions of ultrasonic data. }
\label{fig:4}
\end{figure*}

\textbf{Dynamic rupture propagation:} The horizontal $P$-wave velocity before localised dynamic failure drops throughout the entire sample, from 6.2~km/s down to 5.4~km/s (Figure \ref{fig:4}a). The $V_{\mathrm{P}}$ structure obtained immediately after dynamic failure shows a strong localised low velocity zone ($V_{\mathrm{P}}$ drops by 22\% down to 4.8 km/s) around the fault zone (Figure \ref{fig:4}b). The localised zone of low $V_{\mathrm{P}}$ decreases in width from about 35~mm to 20~mm as the sample is reloaded. $V_{\mathrm{P}}$ recovers throughout the sample during reloading, most notably in the low velocity zone where the minimum $P$-wave velocity increases by 600 m/s to 5.4 km/s (Figure \ref{fig:4}c, d). 

\begin{figure*}
\centering
\includegraphics[scale = 0.9]{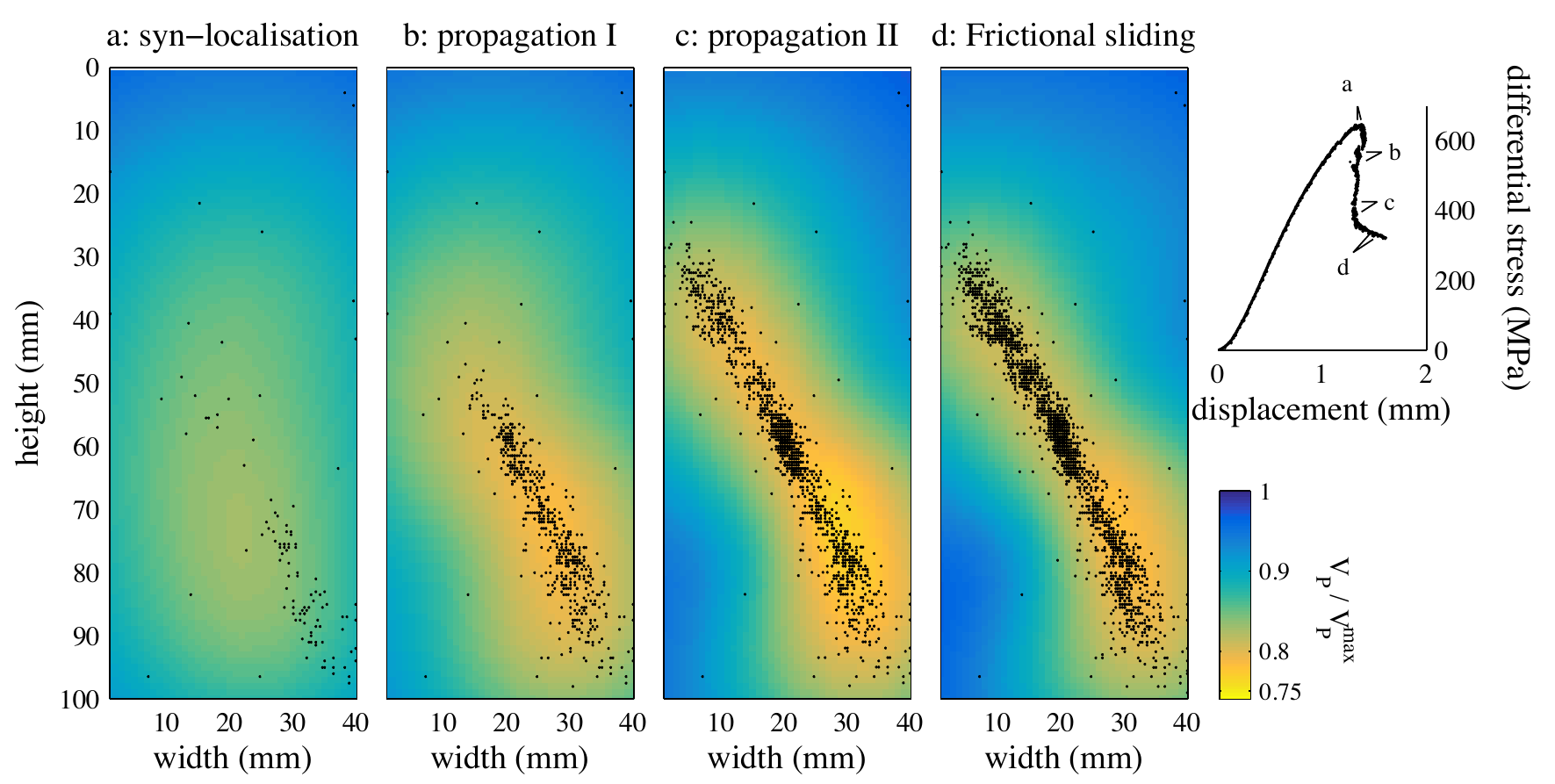}
\caption{Tomographic sections of the horizontal $V_\mathrm{P}$ normalised to the initial velocity through the centre of  sample LN5 (quasi-static rupture propagation), perpendicular to the fault. The four slices show time intervals \textbf{(a)} during localisation of deformation, \textbf{(b)},\textbf{(c)} during two stages of rupture propagation, and \textbf{(d)} during frictional sliding of the fault. The corresponding parts of the stress-displacement curve are indicated on the right. AE source locations up to the time interval shown are projected onto the slice. The AE source locations are within 2.5~mm distance perpendicular to the slide, and were determined using the 3D seismic velocity model. Figure from \citet{aben19}. }
\label{fig:5}
\end{figure*}

\textbf{Quasi-static rupture propagation:} Before the onset of quasi-static rupture, horizontal $V_\mathrm{P}$ decreases from around 6~km/s down to 5~km/s (Figure \ref{fig:5}a). The rupture nucleates near the bottom of the sample and propagates upwards (delineated by the AE source locations), during which a low $P$-wave velocity zone forms around \clearpage\noindent the fault zone (Figure \ref{fig:5}b, c). $V_{\mathrm{P}}$ within the low velocity zone is as low as 4.6~km/s (a 25\% drop relative to unaffected areas outside of the zone). In the wake of the rupture tip, the $P$-wave velocity at some distance from the fault recovers by at most 5\%. After rupture completion at the onset of frictional sliding, $V_\mathrm{P}$ recovers throughout the sample (the minimum $V_{\mathrm{P}}$ rises by about 100 m/s, Figure \ref{fig:5}d).  

\begin{figure*}[t]
\centering
\includegraphics[scale = 0.9]{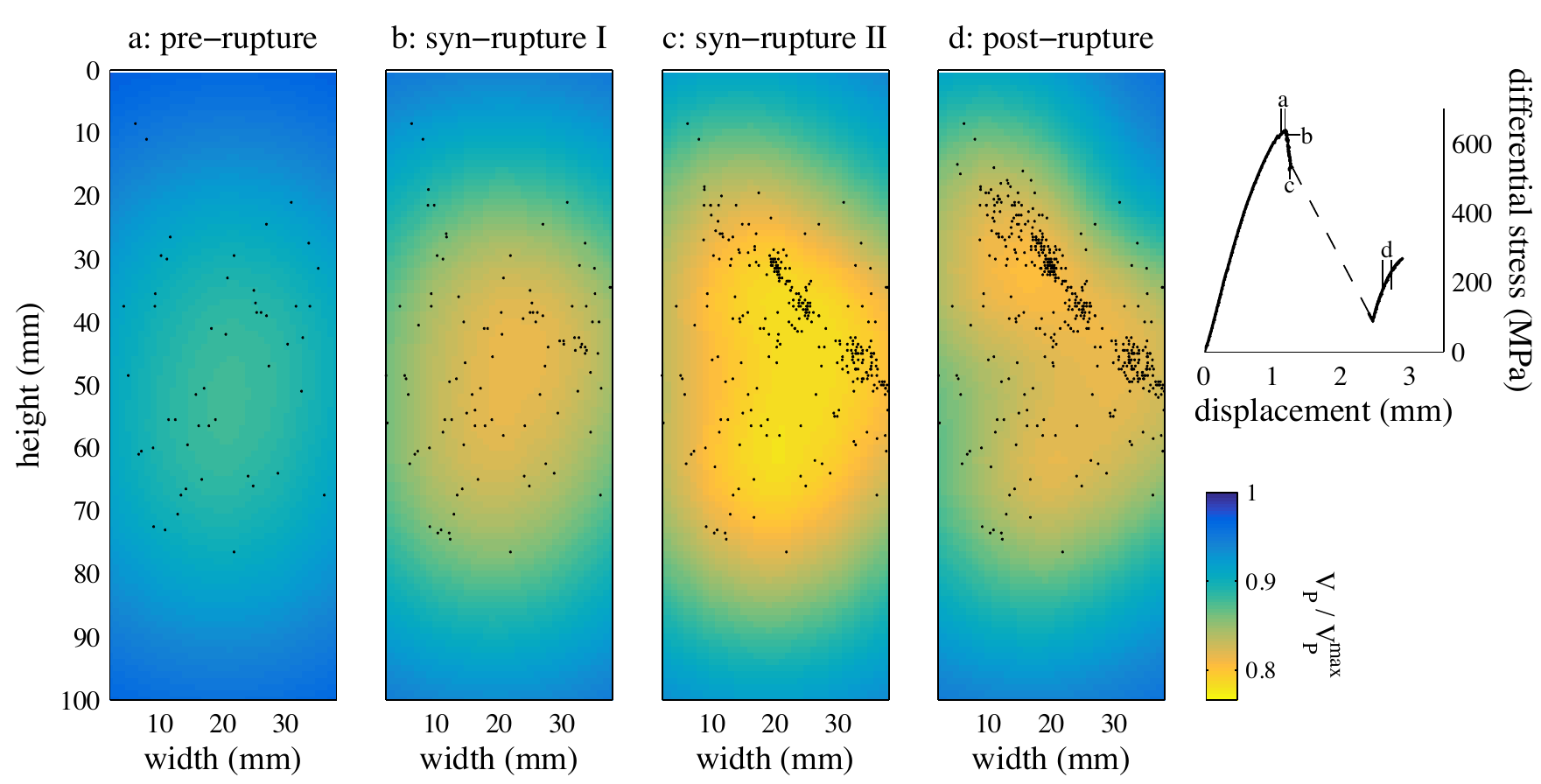}
\caption{Tomographic sections of the horizontal $V_\mathrm{P}$ normalised to the initial velocity through the centre of  sample LN8 (mixed rupture propagation), perpendicular to the fault. The four slices show time intervals \textbf{(a)} during localisation of deformation, \textbf{(b)} and \textbf{(c)} during two stages of rupture propagation, and \textbf{(d)} during reloading of the fault. The corresponding parts of the stress-displacement curve are indicated on the right. AE source locations up to the time interval shown are projected onto the slice. The AE source locations are within 2.5~mm distance perpendicular to the slide, and were determined using the 3D seismic velocity model. }
\label{fig:6}
\end{figure*}

\begin{figure*}
\centering
\includegraphics[scale = 0.9]{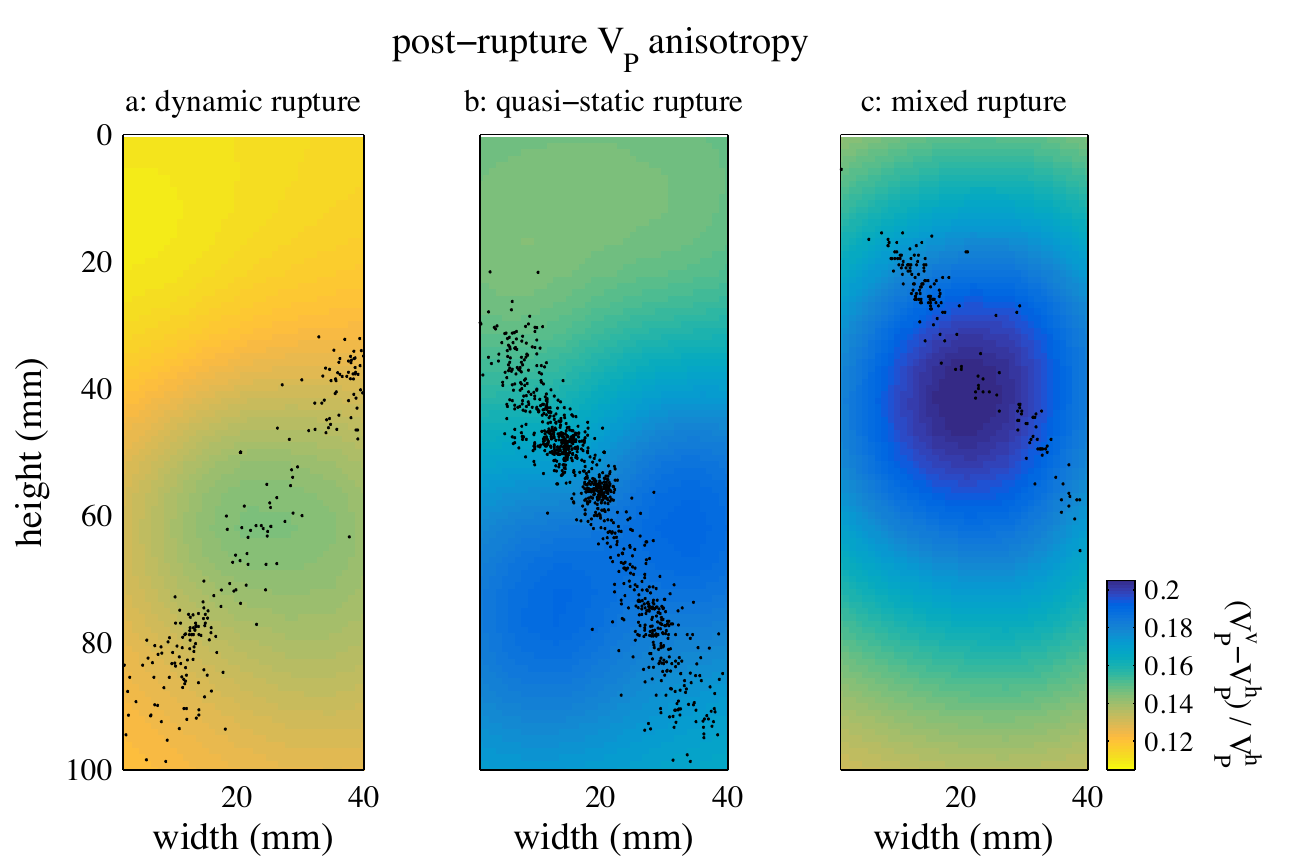}
\caption{$P$-wave anisotropy tomography sections through the centre of the samples and oriented perpendicular to the failure planes, based on ultrasonic data measured during reloading of a dynamically failed sample \textbf{(a)}, after quasi-static rupture completion \textbf{(b)}, and during reloading of a mixed failure experiment \textbf{(c)}. The tomography sections in (a), (b), and (c) are from the same time intervals as the horizontal $P$-wave results in Figure \ref{fig:4}d, Figure \ref{fig:5}d, and Figure \ref{fig:6}d, respectively. AE source locations within 5~mm perpendicular to the section are projected on the section, and only post-rupture AE events are shown.}
\label{fig:7}
\end{figure*}

\textbf{Mixed rupture propagation:} The horizontal $V_\mathrm{P}$ before failure decreases throughout the sample (Figure \ref{fig:6}a), similar to the velocity drop observed before dynamic failure and quasi-static rupture (Figures \ref{fig:4}a and \ref{fig:5}a). The deformation history of this particular sample becomes somewhat complicated after the peak stress: We observe a faint localisation zone, delineated by AE source locations and visible in a polished section (Figure d), that is oblique to the final failure surface. The $V_\mathrm{P}$ structures of the first few time intervals after the peak stress show a low velocity zone around this aborted nascent rupture plane (Figure \ref{fig:6}b). The eventual fault forms after a 50~MPa drop relative to the peak stress, and is embedded in a zone with velocities as low as 4.6 km/s, from an initial velocity of 5.9 km/s -- a 22\% drop (Figure \ref{fig:6}c). The rupture was allowed to propagate dynamically at about 520~MPa. The $V_\mathrm{P}$ structure after failure shows two elongated low velocity zones, one around the main fault zone and one around the `failed' fault zone with velocity reductions of 20\% and 18\%, respectively, a recovery by 200 m/s up to 4.8 km/s (Figure \ref{fig:6}d). 

A localised low $P$-wave velocity zone is observed for the three rupture types (quasi-static, dynamic, and mixed). The minimum horizontal velocities within these zones are of similar order of magnitude: 4.6--4.8 km/s, equal to a 22--25\% drop relative to the initial $P$-wave velocity. These velocity drops are in accordance with the largest drops observed in the path-averaged horizontal $V_\mathrm{P}$ (Figure \ref{fig:3}). The largest velocity decrease for quasi-static and mixed rupture is observed during the propagation of the rupture itself (Figures \ref{fig:4}c and \ref{fig:6}c). For dynamic rupture, the lowest velocities were observed directly after failure and thus provide only an upper bound for the lowest horizontal $P$-wave velocities during dynamic rupture.

\subsubsection{$P$-wave tomography: Anisotropy}
During axial loading up to failure, $V_\mathrm{P}$ anisotropy in dynamic and mixed rupture samples is fairly homogeneous and varies between 10--11\% (i.e., the vertical $P$-wave velocity is 10--11\% higher than the horizontal $P$-wave speed). Some variation in anisotropy near the sample extremities may be caused by lateral confinement from the coupling with the loading column. The anisotropy for the quasi-statically ruptured sample is somewhat higher at 13--15\%, although a 1--2\% variance within the sample is similar to the dynamic and mixed rupture samples.

The $V_\mathrm{P}$ anisotropy adjacent to the ruptured zone increases up to 20\% during quasi-static rupture. The anisotropy outside the ruptured zone remains at 15\%, similar to the pre-rupture anisotropy. Anisotropy measured in the mixed rupture sample, during the quasi-static rupture interval, increases to 19\% around the ruptured zone, and anisotropy outside the ruptured zone remains more or less constant at 12\%. Thus, during rupture the vertical $P$-wave velocity decreases less relative to the horizontal $P$-wave velocity. 

The lowest anisotropy after rupture completion (at residual shear stress) is observed in the dynamically ruptured sample, with a maximum anisotropy of 14\% near the ruptured zone and a 11\% anisotropy outside this zone (Figure \ref{fig:7}a). The maximum anisotropy near the ruptured zone after completion of quasi-static rupture is 19\% (Figure \ref{fig:7}b), which is a small recovery relative to the maximum anisotropy during rupture. The anisotropy in the volume unaffected by rupture remains similar to the pre-rupture anisotropy. The anisotropy after dynamic failure in the mixed rupture sample is 20\% (Figure \ref{fig:7}b), and the minimum anisotropy outside the ruptured zone is 12\%. We thus see in all three rupture experiments an increase in anisotropy around the ruptured zone during and after failure, with the smallest increase after dynamic rupture. We can infer from the horizontal $P$-wave velocity decrease and anisotropy increase in the ruptured zone that the vertical $P$-wave velocity during rupture does not change much. Outside the ruptured zones, the anisotropy during and after rupture remains constant relative to the initial anisotropy just prior to reaching the peak differential stress.

\subsection{Microstructural observations}

\begin{figure*}
\centering
\includegraphics[scale = 1]{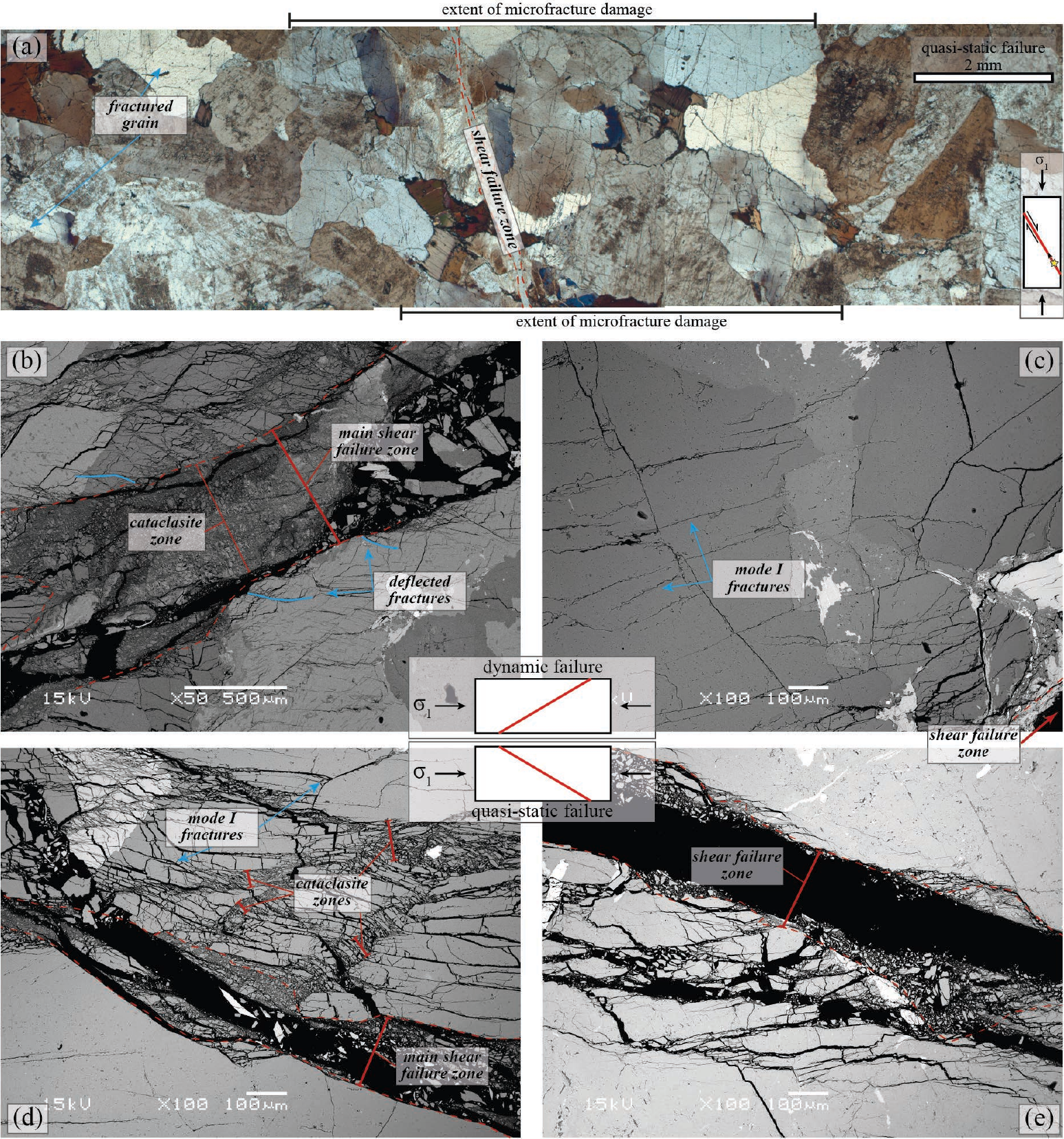}
\caption{\textbf{(a)}: Optical microscope images (transmitted light, cross polarizers at $45^\circ$) perpendicular to the failure zone after quasi-static failure show a damage zone of mm-scale microfracture damage, and heavily fractured individual grains outside this zone. Star and arrow in the inset indicates rupture nucleation and propagation direction. SEM-BSE images of \textbf{(b)},\textbf{(c)}: Dynamically (sample LN7) and \textbf{(d)},\textbf{(e)}: Quasi-statically (sample LN5) failed samples. The principal loading axis is oriented horizontally in all four images. \textbf{(b)}: The main failure plane is delineated with gouge and in some places bound by zones of micro-cataclasite. Surrounding grains show mode I microfractures, of which some deflect into the slip zone. \textbf{(c)}: Off-fault mode I microfracture damage within an individual grain decreases somewhat with distance from the failure plane (bottom right corner). \textbf{(d)},\textbf{(e)}: Secondary shear deformation zones (cataclasite zone) branch of the main failure plane (d), whereas other sections of the main failure plane are relatively straight without secondary structures (e). Off-fault mode I microfracture damage is visible in all cases, but is higher where there is more complexity on the main failure plane. }
\label{fig:8}
\end{figure*}

Study of thin sections by optical microscopy reveals a zone of microfractures of around 1~mm in length and oriented parallel to the loading direction enveloping the shear failure zone (Figure \ref{fig:8}a). For quasi-static shear failure, the extent of this damaged zone is appraised at roughly 8 to 10~mm on the tensile side of the fault, and 2 to 3~mm on the opposite compressive side. Several grains outside this off-fault damage zone have been subjected to extensive fracturing as well (Figure \ref{fig:8}a). We will attempt to quantify our qualitatively assessed order of magnitude damage zone width from fracture density data obtained from SEM images hereafter. First, we describe the microstructures observed at smaller scale in the SEM images, followed by measurements of off-fault microfracture orientation and density.

The SEM images show that the main failure plane resulting from dynamic rupture is surrounded by patches of gouge and cataclasite (Figure \ref{fig:8}b, c), which were not preserved everywhere in the sample during the post-mortem treatment. Whereas the individual particles in patches of gouge cannot be clearly distinguished on the images, the fragments in the cataclasite zones are clearly visible and angular, and show rotation relative to their neighbouring fragments. At 100-500~$\mu$m distance from the main failure zone, the rock contains abundant mode I microfractures oriented parallel to the main loading direction with little to no shear or rotation of fragments (Figure \ref{fig:8}b). Some of these mode I fractures tend to deflect towards the main slip zone (Figure \ref{fig:8}b). Qualitatively, the amount of microfractures decreases with increasing distance from the fault (Figure \ref{fig:8}b, c; Figure \ref{fig:9}a, b), and variation in microfracture density on the scale of individual SEM images is linked to mineral type (for instance, the biotite grain at the top of \ref{fig:9}b is more heavily fractured relative to the feldspar below it). The microstructural damage observed near a quasi-statically formed failure zone is qualitatively similar to that observed after dynamic rupture; patches of gouge and cataclasite zones (Figure \ref{fig:8}d, e) are visible along the main failure zone and some cataclasite zones form secondary brittle shear zones (Figure \ref{fig:8}d). Other parts along the main failure plane are less complex and show only a thin zone of gouge and cataclasite (Figure \ref{fig:8}e). Primarily mode I microfracture damage is observed further away from the fault (Figure \ref{fig:8}d, e). The scope of this study is to quantify off-fault damage related to rupture, and we therefore removed from the traced images the zones of gouge and cataclasite that are clearly related to slip before further analysis of off-fault microfracture damage. Examples of SEM-BSE images with fracture traces are shown in Figure \ref{fig:9}a, b and Figure \ref{fig:10}a, b. 

\textbf{Off-fault microfracture orientations:}
The dominant fracture orientation for off-fault microfractures was obtained from the cumulative length of the major ellipse axis of all the fracture segments in all SEM images that fall within 5$^{\circ}$~intervals measured relative to the loading axis. All intervals are normalised by the interval with the largest cumulative length. The overall dominant off-fault microfracture orientation is parallel to the loading axis for both dynamic rupture (Figures \ref{fig:9}c) and quasi-static rupture (Figure \ref{fig:10}c). The angle of the off-fault microfractures with respect to the fault plane is somewhat larger for the dynamic rupture case relative to the quasi-static one.

\textbf{Off-fault fracture density:}
The surface area of, and fracture traces in, gouge and cataclasite zones and empty fault space has not been used in the calculation of the off-fault microfracture density. The off-fault microfracture density is presented as a function of fault perpendicular distance. We set the origin of each SEM transect (i.e., 0~mm fault perpendicular distance) at the centre of the main failure plane, whose width varies along the fault but remains less than 1~mm (Figure \ref{fig:8}) -- thus none of the SEM images is entirely located in the main failure plane.  

After dynamic rupture, off-fault fracture density, $\rho^\mathrm{frac}$, is around 80~mm/mm$^{2}$ directly adjacent to the failure zone, and drops to 30--40~mm/mm$^{2}$ at about 1~mm distance from the failure zone (Figure \ref{fig:9}d).  Further from the fault, we observe an overall cm scale trend of decreasing $\rho^\mathrm{frac}$ with increasing distance from the fault, superimposed to a mm scale variation. This variation is between 5--30~mm/mm$^{2}$ and also decreases with distance from the fault (Figure \ref{fig:9}d). 

$\rho^\mathrm{frac}$ is around 50~mm/mm$^{2}$ directly adjacent to the quasi-statically formed failure zone (Figure \ref{fig:10}d). Within 1~mm distance, $\rho^\mathrm{frac}$ drops to about 10-20~mm/mm$^{2}$. After this initially steep \clearpage \noindent drop, $\rho^\mathrm{frac}$ decreases to below 10~mm/mm$^{2}$ over 1.5~cm fault perpendicular distance. A mm~scale variation is restricted to about 10~mm/mm$^{2}$ magnitude, and decreases with distance. $\rho^\mathrm{frac}$ measured across the quasi-statically ruptured failure zone is lower and has a lower variance relative to $\rho^\mathrm{frac}$ measured across the dynamically ruptured failure zone.

\subsubsection{Damage zone width}
We shall now attempt to summarise the off-fault microfracture density data in an informative and simple measure. For this, we elect the measure of a damage zone width, allowing a direct comparison with fault damage zones studies in the field and with models that predict the extent of off-fault fracture damage from fault rupture and fault slip. The order of magnitude estimate for damage zone width from optical microscopy will be an independent indicator in determining damage zone widths from microfracture density data.

The fracture damage that was measured in the SEM images was created during three stages, according to AE activity: i) pre-failure microfracturing throughout the volume of the sample during yield, leading to ii) fracture coalescence in the nucleation patch and process zone of the propagating rupture, which is followed by iii) slip-induced damage \citep{tapponnier76}. Here, we are only interested in microfractures formed during stage ii), as it provides a measure for $\Gamma_{\mathrm{off}}$. Although we suppress clearly slip-induced stage iii) damage by removing gouge and cataclasite layers from the traced images, we cannot rule out that some of the off-fault microfracture damage has a slip-related origin.

The damage zone width is defined as the fault perpendicular distance where the fault-related fracture density trend (stage ii) and iii) damage) intersects the background fracture density. We define the background fracture density $\rho^{\textrm{frac}}_{0}$ as the sum of yield-related damage and initial damage already present in the samples prior to the experiment. It is reasonable to assume that stage i) introduced equal amounts of damage in the dynamic and quasi-static failed samples, and so $\rho^{\textrm{frac}}_{0}$ estimated for the quasi-statically ruptured sample represents $\rho^{\textrm{frac}}_{0}$ of the dynamically ruptured sample as well. We note that in the first place, our definition of background fracture damage gives a reference value for $\rho^{\textrm{frac}}_{0}$ for our experiments, and is not, but may approach, the background fracture density as encountered in field studies.

We obtain a measure for the background fracture density $\rho^{\textrm{frac}}_{0}$ for the quasi-static failed sample from the average fracture density of 19 SEM images. The conditions to assume that these SEM images were outside the damage zone were: 1) They are located at 12~mm distance or more from the failure zone, 2) they lack open microfractures more than half the image in length, and 3) they do not contain heavily fractured zones (some example images are shown in SI Text S3). Of these 19 SEM images, the images with the lowest fracture densities (around 2-4 mm/mm$^{2}$) are qualitatively similar to the initial undeformed state of Lanh\'elin granite \citep{siratovich15} -- but $\rho^{\textrm{frac}}_{0}$ based on these images excludes yield-related fracture damage and would give a lower bound only. To obtain a more realistic $\rho^{\textrm{frac}}_{0}$, the 19 SEM images also include some with a higher background fracture damage (up to 10 mm/mm$^2$), but without clear stage ii) or iii) related fracture damage. 

We find that $\rho^{\textrm{frac}}_{0} = 6.2$~mm/mm$^{2}$ with a standard error of 2.8~mm/mm$^{2}$ (Figure \ref{fig:9}d, \ref{fig:10}d). A visual check makes it clear that the trend of decreasing fracture density after quasi-static failure intersects with the background fracture density at around 5 to 10~mm fault parallel distance for most transect (Figure \ref{fig:10}d), which matches the damage zone width first estimated from optical microscopy images (Figure \ref{fig:8}a). Fracture density data after dynamic failure seems to intersect with the established background fracture density at larger fault perpendicular distances between 10 to 20~mm, suggesting a wider damage zone (Figure \ref{fig:9}d). 

The decrease in fracture density with distance may be described by a power law function or exponential function, as is often done for damage zone studies in the field and laboratory \citep[e.g.,][]{faulkner11b, mitchell09, savage11, moore95, ostermeijer20}. The intersection of such a fitted function intersects with the background fracture density threshold provides a damage zone width. This approach applied to a high resolution off-fault damage dataset with a large natural variance results in very large uncertainties on the damage zone width \citep{ostermeijer20}, thus oversimplifying or misrepresenting the actual damage distribution. We nonetheless pursued this approach for each transect and the results are presented in Text S3. For most transects, both on quasi-statically or dynamically failed samples, a power law decay fits best with the data. The damage zone width results are not always sensible and in harmony with our primary observations (Figure \ref{fig:8}a, Figure \ref{fig:9}d, and \ref{fig:10}d) -- for instance, damage zone widths after quasi-static rupture that are much larger than 20~mm (i.e., outside the sample). We therefore use this method merely as an additional guidance, and resolve to manually picking damage zone widths for all transects in both samples. Note that our approach for obtaining damage zone width may differ from that used in field studies: For instance, we may have used a different definition for background fracture density, and we combined additional constraints with the results of fitting a damage decay function. 

The damage zone widths determined for the quasi-statically ruptured sample are between 7 and 13~mm on the tensile side of the fault, and between 4 and 13~mm on the compressional side of the fault (Figure \ref{fig:11}c). These values are in accordance with our simple estimates from optical microscopy. The damage zone width may exceed the measured transect length or the sample width for a number of transects in the dynamically ruptured sample, in which case we ascribe a lower bound value of 20~mm. On both sides of the fault, the damage zone width is between 11 to 20~mm (Figure \ref{fig:11}c). The damage zone after dynamic failure is thus wider by about a factor of two relative to the damage zone created during quasi-static failure. We did not observe a clear trend between damage zone width and distance from rupture nucleation for dynamic and quasi-static rupture (Figure \ref{fig:11}c). The power law exponents of the highest quality power law fits vary between $-0.37$ and $-0.49$ for transects in both samples, these values are similar to those obtained for fault damage decay profiles in crystalline rock in the field \citep{savage11,ostermeijer20}.

\section{Discussion}
\subsection{Ultrasonics and Tomography}
$P$-wave velocity variations during quasi-static and dynamic failure experiments can be ascribed to two effects: 1) Variations in differential stress, where an increase results in closing of pre-existing horizontal microfractures (i.e., perpendicular to the loading axis) and opening of pre-existing vertical microfractures. Closing of horizontal microfractures mostly affects vertical $V_\mathrm{P}$, and opening of vertical microfractures has a greater effect on the horizontal $V_\mathrm{P}$ \citep[][, Chapter 5]{paterson05}. 2) Microfracture formation and growth during quasi-static or dynamic failure reduce the $P$-wave velocity locally. These microfractures are subjected to opening or closing as well. 

\begin{figure*}[h]
\centering
\includegraphics[scale = 0.9]{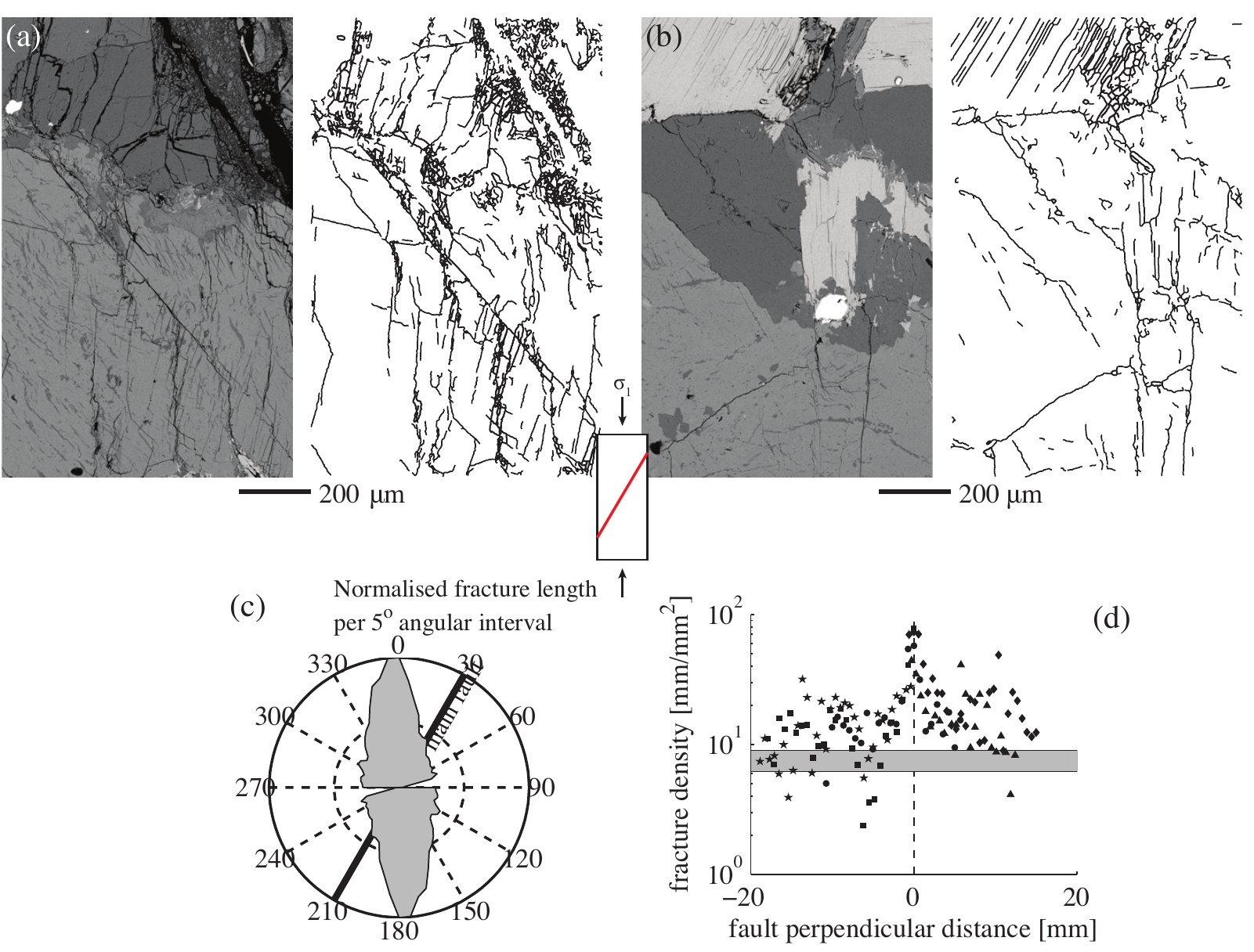}
\caption{Microstructural analysis of a dynamically ruptured sample (sample LN7). \textbf{(a)}: Left panel: SEM-BSE image ($100\times$ magnification) near the main fault. The main fault is visible at the top right, surrounded by abundant fault gouge. Right panel: Fracture traces obtained from the images on the right. \textbf{(b)}: SEM-BSE image (left panel) and fracture traces (right panel) at some distance from the main fault. The locations of (a) and (b) within the sample are indicated in Figure \ref{fig:1}a. \textbf{(c)}: Cumulative fracture length per $5^{\circ}$ angular interval for all SEM images indicates predominantly sub-axial microfracturing. Cumulative length has been normalised by the largest cumulative length. \textbf{(d)}: Fracture density per image as a function of fault-perpendicular distance. The different symbols indicate the different transects defined in Figure \ref{fig:1}a, and the gray area bounds the background fracture density range between the mean and the mean plus one standard error that was established on the quasi-statically ruptured sample LN5. }
\label{fig:9}
\end{figure*}

These two effects are recognised during and after our shear failure experiments, where predominantly vertical microfractures are formed within the damage zone (Figures \ref{fig:9}c and \ref{fig:10}c). These fractures decrease path-averaged $P$-wave velocities along horizontal ray paths more than those along angled ray paths, causing the observed anisotropy (Figure \ref{fig:3}). The closing of the vertical microfractures, caused by the syn-failure differential stress drop, results in partial $P$-wave recovery \citep{passelegue18b}. The lowest $V_\mathrm{P}$ along ray paths intersecting the failure zone are thus observed when the contribution of fracture opening in the rupture process zone dominates over the contribution of fracture closure due to decreasing differential stress (Figure \ref{fig:3}a, asterisks).  We see a similar evolution in $V_\mathrm{P}$ in the time-resolved 3D $P$-wave structure during the stress drop associated to quasi-static failure (Figure \ref{fig:5}b, c): A recovery of $V_\mathrm{P}$ throughout the sample, except near the rupture front where $V_\mathrm{P}$ decreases. \clearpage

\small
After dynamic failure, at the onset of reloading, the path-averaged $V_\mathrm{P}$ along most ray paths rises by a few percent. This is followed by a slight decrease for the remainder of the reloading interval (Figure \ref{fig:3}b). In the 3D velocity structure, the post-rupture $V_\mathrm{P}$ initially increases, in particular within the low velocity zone (Figure \ref{fig:4}b, c). $V_\mathrm{P}$ further decreases with progressive reloading (Figure \ref{fig:4}c, d). This suggests that horizontal microfractures are closed immediately after the dynamic stress drop, resulting in a $P$-wave velocity increase. After horizontal microfracture closure, the opening of vertical microfractures dominates and horizontal $P$-wave velocity decreases. Such a progression is typically observed at the onset of loading of crystalline rock \citep[][, Chapter 5]{paterson05}.

The path-averaged $P$-wave velocities after quasi-static failure are very similar in magnitude to the the $P$-wave velocity measured along the same ray paths after dynamic failure and reloading (Figure \ref{fig:3}). The $V_\mathrm{P}$ drop during reloading of the dynamically failed sample is a near perfect extension of the $V_\mathrm{P}$ increase during the transition to frictional sliding -- the $V_\mathrm{P}$-stress curves of each pair of matching ray paths can be connected fairly well, except for ray path E. This horizontal ray path crosses the fault zone and shows a much larger velocity drop near 300~MPa differential stress in the dynamic case than it does near the same differential stress for the quasi-static case. This may reflect the wider damage zone that was created during dynamic rupture, as this wave path is more sensitive to vertically oriented micro fracture damage than the other wave paths that intersect the fault. 

\subsection{Relationship between microfracture damage and physical properties}
$P$-wave velocity and anisotropy changes are a direct result from changes in the effective elastic moduli of the material. Under the assumption that effective elastic moduli changes are primarily induced by the formation of microfractures, the $P$-wave tomography data contain information about the microfracture density. We use an effective medium approach to relate the seismic velocity from the tomographic data to effective elastic moduli, and obtain a fracture density tensor from these effective elastic moduli. We then compare the obtained microfracture density tensor with the microfracture densities measured on thin sections. 

\subsubsection{Fracture density computation following an effective medium approach}
We adopt the effective medium approach by \citet{sayers95} for a solid containing non-interacting penny-shaped cracks. The overall strain in a cracked solid is the sum of the elastic strain in the matrix (i.e., the constituent minerals of the rock) and the additional strain due to the presence of cracks:
\begin{linenomath*}
\begin{equation}
\varepsilon_{ij} = S^{0}_{ijkl}\sigma_{kl} + \Delta S_{ijkl}\sigma_{kl}, 
\end{equation}
where is $S^{0}_{ijkl}$ the elastic compliance tensor of the matrix, $\sigma_{kl}$ is the stress tensor, and $\Delta S_{ijkl}$ is the change in elastic compliance resulting from cracks given by \citep{sayers95}:
\begin{equation}
\Delta S_{ijkl} = \frac{1}{4}(\delta_{ik}\alpha_{jl} + \delta_{il}\alpha_{jk} + \delta_{jk}\alpha_{il}+\delta_{jl}\alpha_{ik})+\beta_{ijkl}.
\end{equation}
Here, $\delta_{ij}$ is the Kronecker delta, and
\begin{equation}
\alpha_{ij} = \frac{32(1-\nu_{0}^{2})}{3E_{0}(2-\nu_{0})} \frac{1}{V} \sum_{r} (a_{r})^{3} n_{i}^{r} n_{j}^{r}
\end{equation}
\end{linenomath*}
is the second rank crack density tensor for $r$ penny-shaped cracks with radii $a_{r}$ and unit normal vector $n_{i}^{r}$ in a volume of rock $V$ with intact matrix elastic parameters $E_{0}$ and $\nu_{0}$ (Young's modulus and poissons ratio, respectively). $\beta_{ijkl}$ is a fourth rank crack density tensor, the contribution of which can be neglected in case of a dry rock with low poissons ratio \citep{sayers95}.

\begin{figure*}
\centering
\includegraphics[scale = 0.9]{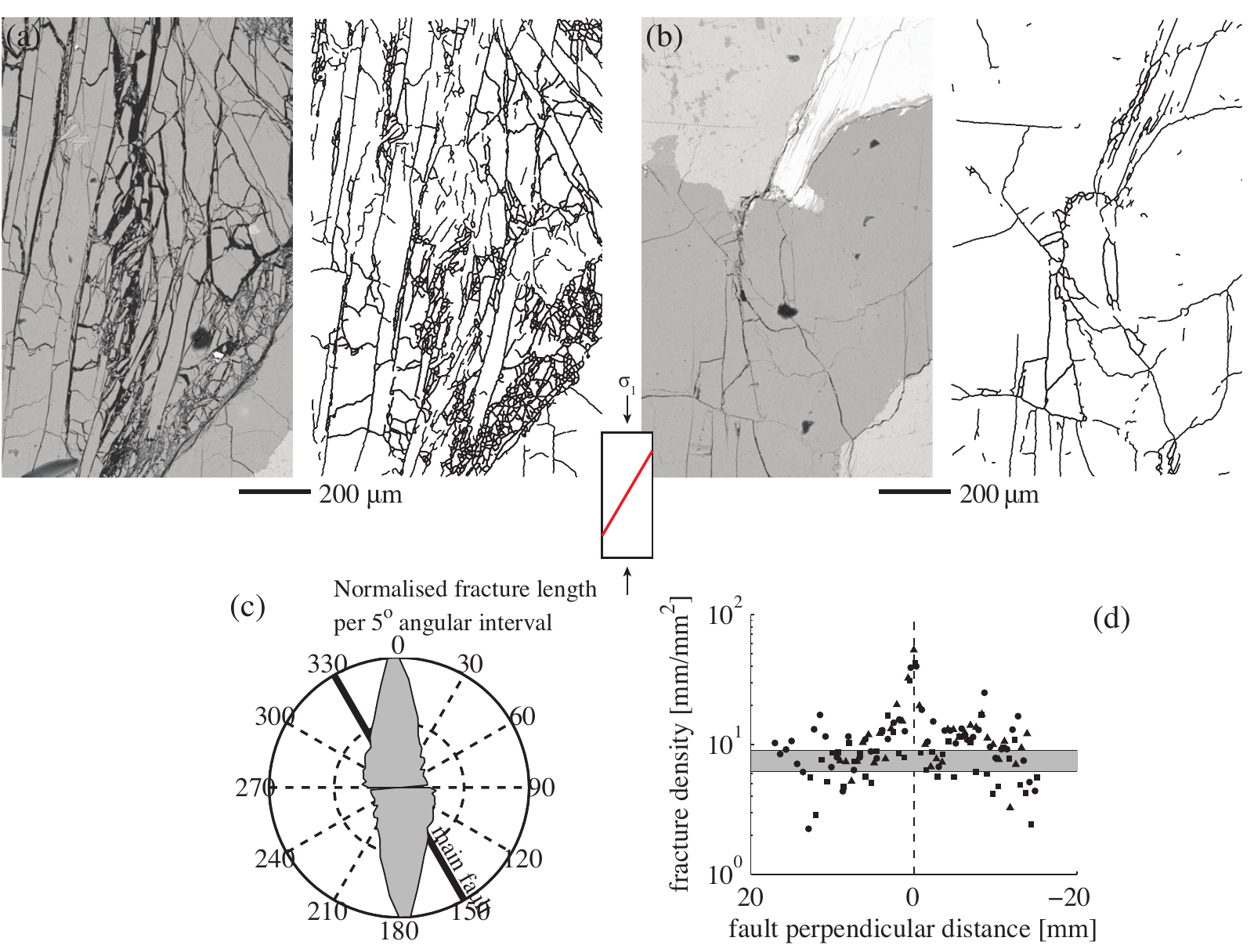}
\caption{Microstructural analysis of a quasi-statically ruptured sample (sample LN5). \textbf{(a)}: Left panel: SEM-BSE image ($100\times$ magnification) near the main fault. The main fault is visible at the bottom right, surrounded by abundant fault gouge. Right panel: Fracture traces obtained from the images on the right. \textbf{(b)}: SEM-BSE image (left panel) and fracture traces (right panel) at some distance from the main fault. The locations of (a) and (b) within the sample are indicated in Figure \ref{fig:1}a. \textbf{(c)}: Cumulative fracture length per $5^{\circ}$ angular interval for all SEM images indicates predominantly sub-axial microfracturing. Cumulative length has been normalised by the largest cumulative length. \textbf{(d)}: Fracture density per image as a function of fault-perpendicular distance. The different symbols indicate the different transects defined in Figure \ref{fig:1}a, and the gray area bounds the background fracture density range between the mean and the mean plus one standard error. }
\label{fig:10}
\end{figure*} 

The tomographic inversion algorithm for the $P$-wave velocity allows for a vertical transverse isotropy in each voxel, which would result from a transversely isotropic orientation distribution of cracks. The orientation distribution of off-fault fracture segments observed across the quasi-static and dynamically formed failure zones (Figures \ref{fig:9}c and \ref{fig:10}c) shows a dominant orientation that is near-vertical to the loading axis. We did not measure fracture orientations parallel to the main failure zone, but we assume these are similar to those measured perpendicular to the failure zone so that the microstructural data is consistent with the tomographic models. For a vertical transverse isotropic distribution of cracks, $\alpha_{11} = \alpha_{22}$ are the horizontal components and $\alpha_{33}$ the vertical component of the crack density tensor. \citet{sayers95} gives the elastic stiffness tensor $C_{ijkl}$ for vertical transverse isotropy in Voigt notation as:
\begin{linenomath*}
\begin{equation}
\begin{aligned}
C_{11} + C_{12} &= (S_{11}^{0} + \alpha_{33})/D, \\
C_{11} - C_{12} &= 1/(S_{11}^{0} - S_{12}^{0} + \alpha_{11}), \\
C_{33} &= (S^{0}_{11} + S^{0}_{12} + \alpha_{11})/D, \\
C_{44} &= 1/(2S_{11}^{0} -2S_{12}^{0} + \alpha_{11} + \alpha_{33}), \\
C_{13} &= -(S_{12}^{0})/D, \\
C_{66} &= 1/(2S_{11}^{0} - 2S_{12}^{0} + 2\alpha_{11}), \\
& \quad D = (S_{11}^{0} + \alpha_{33}) (S_{11}^{0} + S_{12}^{0} + \alpha_{11}) - 2(S_{12}^{0})^{2} .
\end{aligned}
\end{equation}
\end{linenomath*}

Within the frame of reference of the sample, the vertical and horizontal crack densities are:
\begin{linenomath*}
\begin{equation}
\begin{aligned}
\rho_\mathrm{v} &= \frac{2\alpha_{11}}{h}, \\
\rho_\mathrm{h} &= \frac{\alpha_{33}}{h}, \\
& \quad h = \frac{32(1-\nu_{0}^{2})}{3E_{0}(2-\nu_{0})}.
\end{aligned}
\end{equation}
\end{linenomath*}
To find values for $\rho_\mathrm{v}$ and $\rho_\mathrm{h}$, we use a similar inversion protocol to \citep{brantut15b} where we calculate the theoretical compliance tensor $C_{ij}$ for a range of possible values of $\rho_\mathrm{v}$ and $\rho_\mathrm{h}$. From $C_{ij}$, we obtain synthetic values for $V_{\textrm{P}}^{\textrm{v}}$ and $V_{\textrm{P}}^{\textrm{h}}$:
\begin{linenomath*}
\begin{equation} 
\begin{aligned}
V_{\textrm{P}}(\theta) &= \frac{C_{11}\sin{^{2}(\theta)}  +C_{33}\cos{^{2}(\theta)}+C_{44} +\sqrt{M} }{2\rho} \\
& \quad M = \left[ (C_{11} - C_{44})\sin{^{2}(\theta)} - (C_{33}-C_{44})\cos{^{2}(\theta)} \right]^{2} + \left[(C_{13} + C_{44}) \sin{(2\theta)} \right]^{2}, 
\end{aligned}
\end{equation}
\end{linenomath*}
where $\rho$ is the density of the intact rock matrix and $\theta$ the angle with respect to the loading axis, which is $0^\circ$ for $V_{\textrm{P}}^{\textrm{v}}$ and $90^\circ$ for $V_{\textrm{P}}^{\textrm{h}}$. We then use a least absolute criterion to obtain the best fit between synthetic $V_{\textrm{P}}$ and measured $V_{\textrm{P}}$, assuming a Laplacian probability density function, so that we obtain the most likely values for $\rho_\mathrm{v}$ and $\rho_\mathrm{h}$ \citep{tarantola05, brantut11b}. For this, we assume an uncertainty on the measured $V_{\textrm{P}}$ of 200~ms$^{-1}$.

The $P$-wave tomography models obtained after quasi-static, dynamic, and mixed failure provide the observed values for $V_{\textrm{P}}^{\textrm{h}}$ and $V_{\textrm{P}}^{\textrm{v}}$. We took these velocities along a fault-perpendicular transect through the centre of each sample. $S^{0}_{ij}$ and $h$ were calculated from $\nu_{0} = 0.20$ and $E_{0}$. We used a value for $E_{0}$ derived from the path-averaged $V_{\textrm{P}}$ measured at peak stress, and take $\rho = 2660$~kg~m$^{-3}$.

\begin{figure*}
\centering
\includegraphics[scale = 0.9]{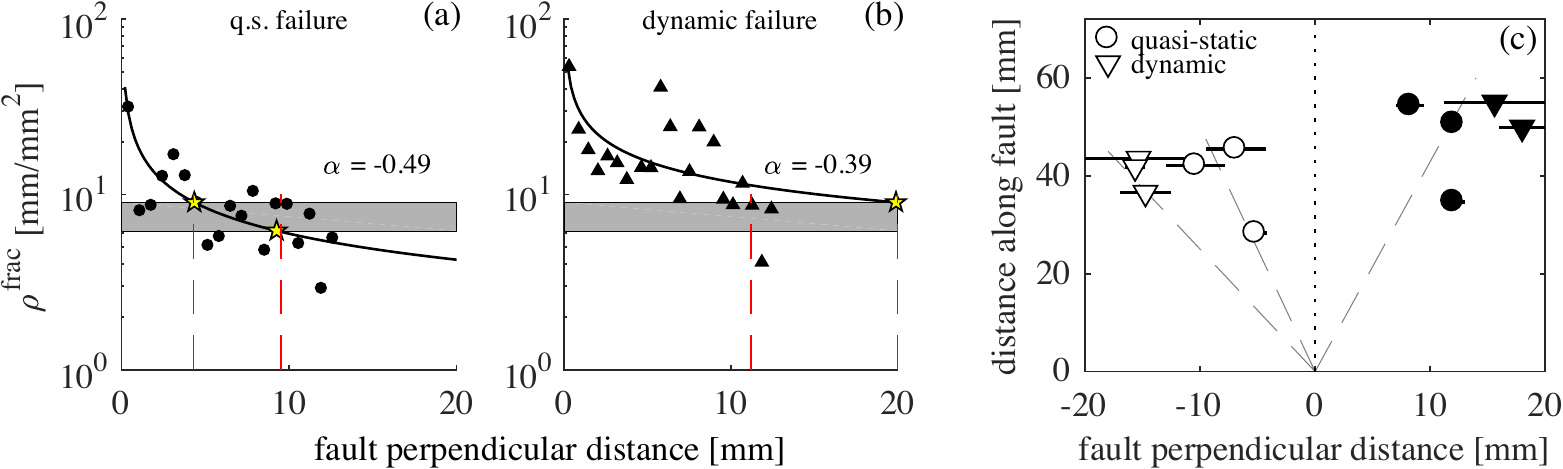}
\caption{Fracture density along a single fault perpendicular transect after quasi-static \textbf{(a)} and dynamic \textbf{(b)} failure, fitted with a power law function (exponent $\alpha$ given for both fits). Yellow stars indicate the intersection of the fitted function with the gray area, which bounds the background fracture density range between the mean and the mean plus one standard error. The manually picked range for the damage zone width is shown by the red lines (upper and lower bound). See SI Text S3 for all transects. \textbf{(c)}: Distance along the fault versus damage zone width, for transects in samples subjected to quasi-static (circles) and dynamic (triangles) loading. White symbols are transects on the tensile side of the fault, black symbols on the compressional side of the fault (see Figure \ref{fig:1}a, b). Distance along the fault is measured in the direction of rupture propagation. Dashed lines indicate the approximate trend. }
\label{fig:11}
\end{figure*}

Computed vertical crack densities after dynamic failure increase from $\rho_{\mathrm{v}} = 0.09-0.10$ at the edge of the sample to $\rho_{\mathrm{v}} = 0.18$ near the failure zone (Figure \ref{fig:12}a). The horizontal crack density $\rho_{\mathrm{h}}$ increases from 0~near the edge of the sample to about 0.04 near the failure zone. These increasing crack densities suggest a damage zone width of about 20~mm on both sides of the fault, but we exercise caution with this measure as it it is near the resolution of the tomography imposed by the correlation length. After quasi-static failure, computed crack densities near the edge of the sample are somewhat lower compared to those computed for dynamic failure ($\rho_{\mathrm{v}} = 0.04-0.07$~and~$\rho_{\mathrm{h}}$ is negative, Figure \ref{fig:12}b), but show a stronger increase near the failure zone where they show the same peak values ($\rho_{\mathrm{v}} = 0.18$~and~$\rho_{\mathrm{h}} = 0.03$). The negative fracture densities near the edge of the sample result from a slight underestimate of the value for $E_{0}$, which was obtained from path-averaged $V_{\textrm{P}}$ measurements. Such an error is expected in the absolute values of the path-averaged $V_{\textrm{P}}$, but has minor consequences for the change in crack densities. The crack densities after mixed failure are similar to those computed after quasi-static and dynamic failure, but show a strongly asymmetric distribution across the failure zone (Figure \ref{fig:12}c). The higher crack densities on one side of the fault (positive fault-perpendicular distance in Figure \ref{fig:12}c) coincide with the nascent secondary failure zone (Figure \ref{fig:2}d). 

\subsubsection{Off-fault microfracture density from microstructures compared to crack density}
The horizontal and vertical crack densities, $\rho_{\mathrm{h}}$ and $\rho_{\mathrm{v}}$, computed from ultrasonic wave velocities have units of m$^{3}$/m$^{3}$ and are directly derived from the crack tensor components $\alpha_{11}/h$ and $\alpha_{33}/h$ (equation [5]). The off-fault microfracture density $\rho^{\mathrm{frac}}$ obtained from microstructures is measured in m/m$^{2}$ and is a scalar quantity. In order to compare the two methods, we convert the off-fault microfracture traces to tensor components $\alpha_{11}/h$ and $\alpha_{33}/h$. We remain in the spirit of the effective medium approach by assuming that the sample contains a transversely isotropic orientation distribution of penny-shaped fractures so that $\alpha_{11} = \alpha_{22}$ and all cracks have a radius $a_{r}$. We treat each traced fracture segment as a trace through an individual penny-shaped fracture. 

The centre of a traced fracture does not necessarily lie on the SEM image plane, which means that the fracture trace length $t$ is equal to or smaller than the true fracture diameter $2a$. The mean fracture radius $\bar{a}$ was obtained from the mean measured trace length $\bar{t}$ so that $2\bar{a}  =  (\pi/2) \bar{t}$ \citep{oda83}. To determine $\bar{t}$, we first determined the probability distribution of trace lengths. We followed the approach of \citet{rizzo17a} to find the best type of probability distribution that describes the trace lengths in a single SEM image: 1) Maximum likelihood estimators were used to fit power law, exponential, and log-normal distributions, and 2) the goodness-of-fit for all three types of distributions was tested using the Kolmogorov-Smirnoff test, giving a probability for each distribution. We obtained a $>90\%$ probability for a log-normal distribution of trace lengths for a majority of the images, and a power law distribution for the remaining images. For log-normal distributed trace lengths, $\bar{t}$ was calculated from the first moment of the distribution. $\bar{t}$ cannot be determined from the first moment of a power law distribution, and we therefore took the mean of the measured traces. 

The unit vectors in equation [3] for each individual fracture segment are given by $\cos(\theta)$ and $\sin(\theta)$ for the tensor components $\alpha_{11}$ and $\alpha_{33}$, respectively (Figure \ref{fig:1}c). Each image intersects only those fractures that have their centre within $\bar{a}$ distance perpendicular to the image plane \citep{oda83}, under the assumption the average out-of-plane fracture orientations (i.e., rotation with respect to the sample axis) are perpendicular to the image. This assumption agrees with the assumption of a transversely isotropic fracture orientation distribution. The volume $V$ associated to the fracture traces on each SEM image is then given as $V = S\bar{a}$, where $S$ is the surface area of the image. With the parameters $\bar{a}$, $V$, and $\theta$, and equations [3] and [5], we obtain $\rho_{\mathrm{h}}$ and $\rho_{\mathrm{v}}$ for each image. 

\begin{figure*}
\centering
\includegraphics[scale = 0.9]{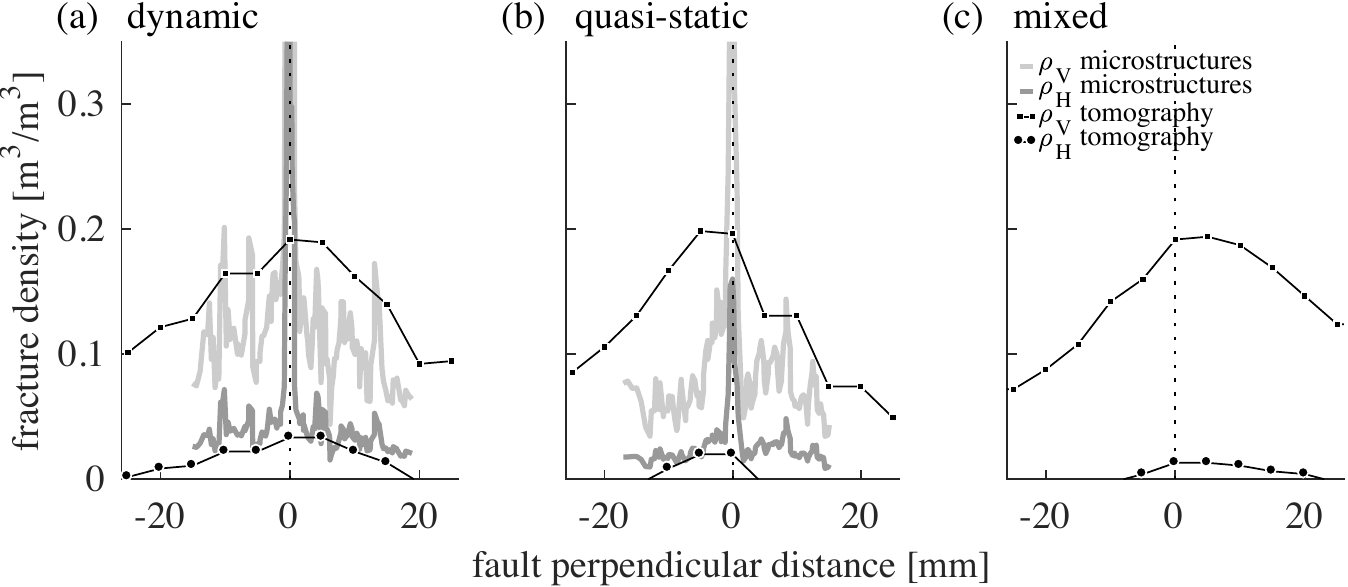}
\caption{Horizontal and vertical fracture densities ($\rho_{\mathrm{h}}$ and $\rho_{\mathrm{v}}$) computed from the $P$-wave tomography models after \textbf{(a)}: Dynamic failure, \textbf{(b)}: Quasi-static failure, and \textbf{(c)}: Mixed failure. The gray curves correspond to horizontal and vertical fracture densities obtained from the off-fault microfracture traces on post-mortem thin section after dynamic and quasi-static failure. }
\label{fig:12}
\end{figure*}

Values for $\rho_{\mathrm{h}}$ and $\rho_{\mathrm{v}}$ obtained from the microstructures agree well with those computed from the tomography models, for both the dynamically and quasi-statically failed samples (Figure \ref{fig:12}a, b), except near the failure zone. Here, between 0 and 2~mm distance from the fault, the microfracture densities from microstructures are up to an order of magnitude higher than those obtained from $V_\mathrm{P}$. This can be ascribed to the difference in spatial resolution of the two methods. Nonetheless, the primary (i.e., cm scale) features of the damage around the failure zone are captured by both direct observation of microfractures and by $P$-wave tomography. 

Our results show that $P$-wave tomography combined with an effective medium theory can quantify localised zones of fracture damage. The use of a normalised fracture density, such as the one presented here, has the advantage of direct applicability with other effective medium models, for instance to predict hydraulic properties \citep{gavrilenko89, gueguen03}. We therefore propose that high resolution geophysical measurements of wave speeds from dense arrays \citep{ben-zion15} combined with microstructural characteristics measured in the field \citep{rempe13, rempe18} or from borehole data \citep{jeppson10} can reveal the physical properties around fault zones. Such data can be used to calibrate the findings of rupture simulations that allow for off-fault energy dissipation \citep{bhat12, thomas18, okubo19}, and can be compared to laboratory failure experiments such as those presented here.

\subsection{Rupture energetics}
We first provide estimates for $\Gamma$ and $W_\mathrm{b}$ for all shear failure experiments from the mechanical stress and strain data. We then show that the damage zone width established after quasi-static and dynamic failure is the results of stresses induced by rupture and not by slip, so that we can calculate $\Gamma_\textrm{off}$ thereafter. We provide estimates for $\Gamma_\textrm{off}$ for quasi-static and dynamic ruptures based on microstructural observations, and discuss the implications. A similar estimate may be obtained for $\Gamma$ dissipated on the fault by quantifying the cumulative fracture surface in gouge and cataclasites that make up the main shear failure zone, but this is beyond the scope of this study. We leave such an endeavour to future studies, as difficulties need to be tackled regarding gouge preservation and resolution limits on identifying the smallest gouge grain sizes.

\subsubsection{Fracture energy and breakdown work from mechanical data}
We calculated breakdown work $W_\mathrm{b}$ by converting the measured stress and axial strain data to shear stress and slip along the failure zone, following the steps described by \citet{wong82, wong86}. The area under the shear stress versus slip curve in excess of the residual shear stress gives a measure for $W_\mathrm{b}$. We measured a residual shear stress of 140~MPa at the end of the quasi-static rupture experiment after 0.83~mm slip (Figure \ref{fig:2}b), whereas the residual shear stress measured after reloading the samples after dynamic and mixed failure was 120~MPa (Figure \ref{fig:2}b). This suggests that the quasi-statically created failure zone had not yet reached its residual frictional strength yet, supported by the convergence of the quasi-static failure stress-strain curve towards this value. Residual shear stresses of 140~MPa and 120~MPa after a slip distance of 0.83~mm give us quasi-static values for $W_\mathrm{b}$ of 37~kJm$^{-2}$ and  53~kJm$^{-2}$, respectively. These are lower bounds for quasi-static $W_\mathrm{b}$, as more slip would have been accrued towards continued weakening down to 120~MPa. 

$\Gamma$ for quasi-static failure may be calculated in the same manner, up to the shear stress and slip distance at which the failure zone through the sample was completed. We thereby assume that all breakdown work done to drop the strength of the failure zone from the peak stress down to shear stress of rupture completion was dissipated to create the failure zone, including formation of off-fault microfractures and gouge associated to the rupture. The rupture was completed and the failure zone fully formed at a shear stress of 155~MPa and a slip distance of 0.44~mm, as established by \citet{aben19} using AE source locations.  From the shear stress and slip data we then obtain $\Gamma = 27$~kJm$^{-2}$ for quasi-static failure \citep{aben19}. 

$W_\mathrm{b}$ cannot be established directly from the mechanical data measured during dynamic failure, as the elastic unloading of the loading column is measured rather than the drop in shear stress of the fault. Order of magnitude estimates for dynamic $W_\mathrm{b}$ and $\Gamma$ may be obtained by approximating the loading system as a simple spring-slider model, similar to \citet{beeler01}, and solve the force balance by assuming some slip-weakening law. We tried this approach, but found results that are too erroneous to be useful. This is most likely because the model assumes a constant piston mass during dynamic failure, which is violated at short failure time scales by inertia of the piston. 

\citet{lockner91} pointed out that it is not strictly correct to use the average shear stress and average slip measured on the sample during failure for calculating $W_\mathrm{b}$ (and $\Gamma$), since the size of the rupture tip process zone is smaller than the sample size. The assumption in using the average shear stress and average slip for the analysis of $W_\mathrm{b}$ \citep{rice80b, wong86} is that the fault is created in the entire sample at the peak stress (i.e., the sample is a point on the trajectory of a propagating rupture) -- which we show is not the case. It is however encouraging that by using this approach, similar order of magnitude values for $W_\mathrm{b}$ have been found on granitic rock samples with different diameters: 16~mm \citep{wong82}, 40~mm \citep{aben19}, and 76~mm \citep{lockner91}. 

\subsubsection{Damage zone width}
Microfracture damage observed after quasi-static and dynamic failure were induced by both rupture and slip. We observe a wider damage zone after dynamic rupture relative to that observed after quasi-static rupture. Part of the slip-related damage was corrected for by removing gouge- and cataclasite patches from the traced images prior to establishing off-fault microfracture densities and damage zone widths (Figure \ref{fig:11}). Nonetheless, some of the remaining off-fault microfractures may be induced by slip rather than rupture. The sample subjected to dynamic failure (sample LN7) accumulated 3.22~mm slip, whereas the sample subjected to quasi-static failure (sample LN5) accumulated 0.83~mm slip. The microstructural record after dynamic failure may thus contain more slip-related off-fault microfracture damage. Before we provide an estimate for $\Gamma_\textrm{off}^\mathrm{surf}$, achieved by combining the damage zone width and microfracture density, we assess whether the difference in damage zone width obtained for quasi-static and dynamic rupture is an effect of rupture velocity or an effect of the difference in accumulated fault slip. 

Off-fault damage during rupture can be caused by stresses around the rupture tip. The geometry of the rupture tip stress field changes with rupture velocity so that damage is created in a larger area around the rupture tip at higher rupture velocity \citep{poliakov02, rice05}, increasing the damage zone width. Slip along a rough fault (i.e., asperities slipping past each other) causes additional stresses in the host rock around the asperities, and these stress heterogeneities result in off-fault damage. With progressive slip along rough faults, progressively larger asperities are dragged past each other and the additional off-fault stresses act over an increasingly larger area \citep{chester00}. The damage zone width is thus also expected to increase with increasing slip. 

The sample subjected to dynamic rupture has experienced both a larger rupture velocity and a larger amount of slip relative to the quasi-statically ruptured sample. Here, we compute the damage zone width as a function of rupture velocity by adopting the analytical solution by \citet{poliakov02} for the elasto-dynamic stress field in a rupture tip process zone for a non-singular slip-weakening rupture. We make the assumption of small scale yielding: The fracture energy is dissipated before the remainder of the breakdown work is done. This means that the initial drop in shear strength in the rock from peak strength down to 155~MPa is solely ascribed to dissipation of $\Gamma$, and further reduction in shear stress is caused by other slip-weakening processes. This assumption seems justified, based on the quasi-static and mixed failure experiments: The initially steep slope in shear stress versus slip during rupture (Figure \ref{fig:2}b) causes stronger stress concentrations relative to the less steep slope of the curve after rupture completion. We therefore expect that the initial steep stress drop determines the damage zone width. We also predict the damage zone width as a function of slip by using the analytical solution by \citet{chester00} for the stress field along a rough frictional fault in an elastic material. Using these two models, and realistic input parameters obtained from the rupture experiments, we then asses which parameter (rupture velocity or slip) is responsible for the observed difference in damage zone width between our quasi-static and dynamic rupture experiments. 

\paragraph{Rupture tip process zone model} 
We consider the 2D case of a mode~II rupture that propagates parallel to the $x$-direction at $z = 0$. The stress in the rupture tip process zone is given by:
\begin{linenomath*}
\begin{equation}
\sigma_{ij} = \sigma^{0}_{ij} + \Delta \sigma_{ij}, 
\end{equation}
where $\sigma^{0}_{ij}$ is the far-field stress state on the sample and $\Delta \sigma_{ij}$ are the additional stress components caused by the rupture that are given by equations [A3] and [A11] in \citet{poliakov02}. To remove the stress singularity at the rupture tip, the shear stress drops linearly from peak stress $\tau_{\mathrm{p}}$ to residual strength $\tau_{\mathrm{r}}$ over a slip-weakening zone of size $R$. For an infinite elastic medium, $R$ decreases in size with increasing rupture velocity $v$ \citep{rice80}:
\begin{equation}
R = \frac{R_{0}}{g(v)} \qquad \textrm{with} \qquad R_{0} = \frac{9 \pi}{32(1-\nu)} \frac{\mu \, \delta_\mathrm{c}}{(\tau_{\mathrm{p}} - \tau_{\mathrm{r}})}, 
\end{equation}
\end{linenomath*}
where $R_{0}$ is the quasi-static limit of $R$, and $\mu$ and $\nu$ are the shear modulus and poissons ratio of the host rock respectively. The function $g$ depends on $v$, and on the $P$- and $S$-wave velocities of the material \citep{poliakov02}. The model parameters were obtained from the mechanical and ultrasonic data measured during the quasi-static rupture experiment on sample LN5 (Table \ref{tab:2}), where we calculated $\mu$ from $E$ and $\nu$. These parameters yield $R_0 = 0.14$~m ,and $R$ decreases down to 0.02~m at $v = 0.9 \times c_\mathrm{s}$. The normal stress on the fault $\sigma_{zz}^{0}$, stress ratio $k = \sigma_{xx}^{0} / \sigma_{zz}^{0}$, and $\tau_{\mathrm{p}}$ were calculated from the mechanical data at the onset of rupture. The residual shear stress $\tau_{\mathrm{r}} = 155$~MPa after $\delta_\mathrm{c}  = 0.44$~mm slip at rupture completion. Note that in the experiments, the shear stress along the failure zone drops further from 155~MPa to 120~MPa, but this occurs after rupture completion and any off-fault damage accrued during this stress drop is not part of off-fault damage related to the rupture. 

\paragraph{Rough fault model} 
We consider the 2D case of a strike-slip fault parallel to the $x$-direction at $z = 0$. We consider a uniform displacement $U$ along a fault with coefficient of friction $\mu$ that has a sinusoidal perturbation:
\begin{linenomath*}
\begin{equation}
A\sin(2\pi x/L),
\end{equation}
where $L$ is the wavelength and $A = \gamma L / 2\pi$ is the amplitude of the perturbation. $\gamma$ is a dimensionless roughness factor. The stress around the rough or wavy fault is given by:
\begin{equation}
\sigma_{ij} = \sigma^{0}_{ij} + \Delta \sigma_{ij}, 
\end{equation} 
where the stress perturbations caused by the sinusoidal fault $\Delta \sigma_{ij}$ are (equations [10a-10c] in \citet{chester00}):
\begin{equation} 
\begin{aligned}
\Delta \sigma_{xx} &= \exp(-l z) \left[(-1+lz) \cos{(lx)} + f(2-lz) \sin{(lx)} \right] B, \\
\Delta \sigma_{zz} &= \exp(-l z) \left[(-1-lz) \cos{(lx)} + (f lz) \sin{(lx)} \right] B, \\
\Delta \sigma_{xz} &= \exp(-l z) \left[f(-1+lz) \cos{(lx)} + (lz) \sin{(lx)} \right] B, \\
\end{aligned}
\end{equation}
with $l = 2\pi /L$ and
\begin{equation}
B  = \frac{\pi E \delta \gamma}{4(1-\nu^{2})}.
\end{equation}
\end{linenomath*}
The normal stress on the fault $\sigma_{zz}^{0}$, stress ratio $k = \sigma_{xx}^{0} / \sigma_{zz}^{0}$, and coefficient of friction $f$ were calculated from the mechanical data at the onset of frictional sliding in the quasi-static experiment (Table \ref{tab:2}). The surface roughness was estimated from a post-mortem cross section perpendicular to the fault plane (Figure \ref{fig:2}c), where the main fault interface (length of the order of 100~mm) shows a waviness in the order of 1~mm, giving $\gamma = 10^{-2}$. During rupture the elastic constants around the fault interface drop by around 50\%, as observed in the $P$-wave tomography results during and after rupture (Figure \ref{fig:4}, \ref{fig:5}, and \ref{fig:6}). To take into account this elastic weakening by the rupture process zone prior to significant slip, Youngs modulus $E$ is half that of the intact rock. 

\begin{table}
\centering
{
\begin{tabular}{r | c c}
 & Rupture model & Roughness model\\ \hline
Young's modulus $E$ & 88 GPa & 44 GPa \\
poissons ratio $\nu$ & 0.22 & 0.22 \\
normal stress $\sigma_{\mathrm{zz}}^{0}$ & 262 MPa & 180 MPa \\
stress ratio $k$ & 2.2 & 1.8 \\
fracture energy $\Gamma$ & $27\times10^{3}$ & - \\
peak shear stress $\tau_{\mathrm{p}}$ & 280 MPa & - \\
residual shear stress $\tau_{\mathrm{r}}$ & 155 MPa & - \\
surface roughness $\gamma$ & - & $10^{-2}$ \\
coefficient of friction $f$ & - & 0.75 
\end{tabular}}
\caption{Model parameters for the rupture tip process zone model and rough fault model. }
\label{tab:2}
\end{table}

\paragraph{Failure criterion and damage zone width}
The 2D off-fault stress tensor around a rupture tip was calculated for a range of rupture velocities from $10^{-6} \times V_{\textrm{S}}$ (i.e., quasi-static rupture) up to $0.9 \times V_{\textrm{S}}$ (i.e., rupture velocity near the Rayleigh wave speed). The stress tensor for a wavy fault was calculated for a range of slip distances between 1~mm and 10~mm, for a range of perturbation wavelengths between 0.2~mm and 150~mm. We used a coulomb failure criterion to assess the damage zone width that can be expected from both mechanisms, where the maximum shear stress $\tau^{\mathrm{max}}$ and coulomb shear stress $\tau^{\mathrm{coulomb}}$ are defined as:
\begin{linenomath*}
\begin{equation}
\begin{aligned}
\tau^{\mathrm{max}} &= \sqrt{(\sigma_{xx} - \sigma_{zz})^{2} /4 + \sigma_{xz}^{2}} \\
\tau^{\mathrm{coulomb}} &= (\sigma_{xx} - \sigma_{zz}) \sin{(\phi)}/2 \\
\end{aligned}
\end{equation}
\end{linenomath*}
where $\phi = \tan^{-1}{(f^{\mathrm{DZ}})}$, and $f^{\mathrm{DZ}}$ is the coefficient of friction within the damage zone, for which we take $\mu^{\mathrm{DZ}} = 1$ to represent intact rock \citep{chester00}. We expect damage where $\tau^{\mathrm{max}} / \tau^{\mathrm{coulomb}} > 1$ or $\tau^{\mathrm{max}} / \tau^{\mathrm{coulomb}} < 0$. The damage zone width is the largest fault parallel distance at which the failure criterion is satisfied. 

\begin{figure}
\centering
\includegraphics[scale = 0.8]{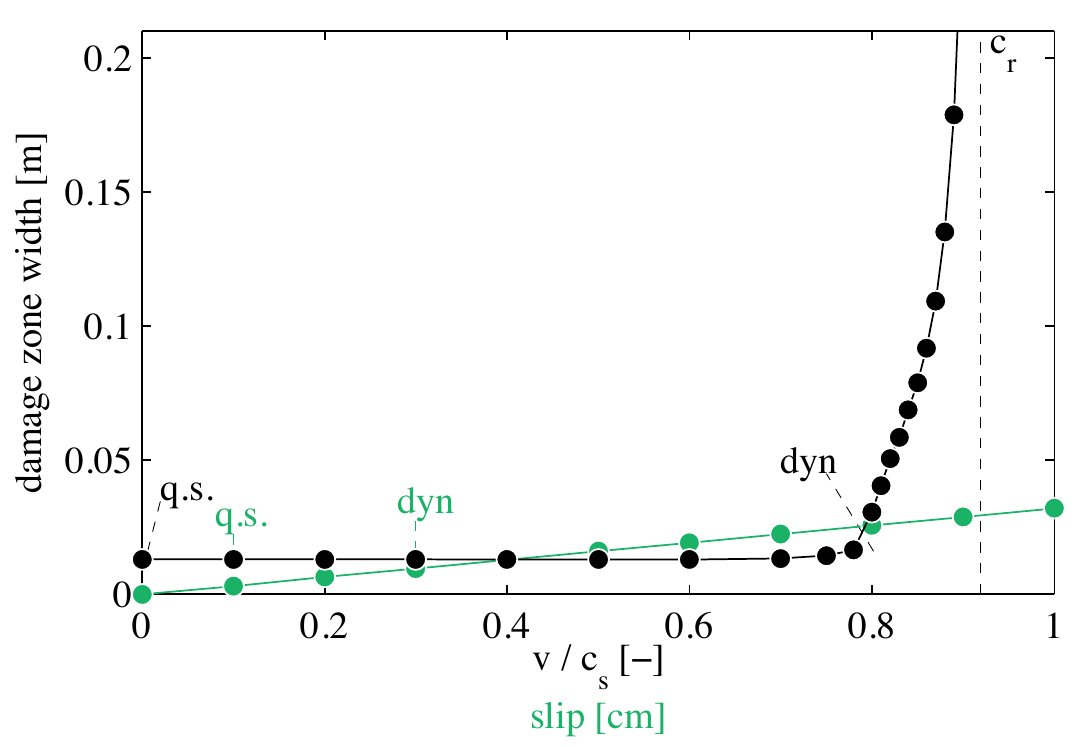}
\caption{Damage zone width as a function of normalised rupture velocity (black) and fault slip (gray), based on the stresses around a propagating rupture tip \citep{poliakov02} and the stresses caused by a wavy perturbation of a frictional interface \citep{chester00}. The rupture velocity of the quasi-static rupture, and the total slip of the quasi-static and dynamic rupture experiments are highlighted. The approximate rupture velocity of the dynamic rupture experiment is indicated where the damage zone width is double that off the quasi-static rupture, in accordance with the microstructural results. }
\label{fig:13}
\end{figure}

The expected fault damage zone width for ruptures propagating below the Rayleigh wave speed increases from about 8~mm at normalised rupture velocities between 0 and 0.75 up to 200~mm or more as the rupture velocity approaches the Rayleigh wave speed (Figure \ref{fig:13}). The damage zone width predicted for increasing slip along a rough fault shows a linear increase with slip (Figure \ref{fig:13}). The damage zone width observed after quasi-static rupture is around 10~mm (Figure \ref{fig:11}c), which is similar to that predicted for a low velocity rupture, whereas the damage zone width predicted to result from slip is less then 3~mm for 0.8~mm of slip accumulated during the experiment. Dynamic rupture resulted in 3~mm total slip, giving a damage zone width of 8.6~mm according to the wavy fault model (Figure \ref{fig:13}). This does not match with the observed damage zone width of around 15 to 20~mm (Figure \ref{fig:11}c). Although the rupture velocity for this experiment was not measured, a damage zone width of 20~mm can result from a dynamic rupture velocity of about $0.8\times V_{\textrm{S}}$.

These results are first order estimates only: The parameters for the rupture model were taken from the quasi-static rupture data, whereas for the dynamic rupture the stress drop may be larger, the coefficient of friction of the fault may be lower, and the breakdown work larger. The rupture model strictly applies to an infinite medium, which may explain why the calculated process zone size $R_0$ is larger than the actual sample size. For the stresses resulting from slip on a rough fault, the background stresses are assumed to be constant, whereas in our experiments initial slip is accumulated within the rupture process zone where the shear stress and the coefficient of friction are not constant. However, this slip-weakening distance for the quasi-static case is less than 1~mm, after which the applied stresses in the experiment remain more or less constant. The simulated off-fault damage at lower velocity ruptures ($<0.7$ normalised velocity) is mostly on the tensile side of the fault plane, similar to the results of \citet{poliakov02}. Other simulations for rupture-induced off-fault damage also predict a strong asymmetry in off-fault damage distribution, with most damage occurring on the tensile side of the fault \citep[e.g.,][]{rice05, xu15, thomas18}. In our experiments however, fracture damage occurs in equal amounts on both sides of the fault, and some microstructural studies on a off-fault damage surrounding an experimentally formed shear fracture did not observe a clear damage asymmetry either \citep{moore95, zang00}. This may be due to several reasons: 1) Rupture simulations are performed in a large continuum with constant far-field stresses, whereas our experiments were performed on a 100~mm by 40~mm cylinder where boundary effects may alter the off-fault stress fields as described by the models. 2) The orientation of the principal stresses (i.e., stress ratio $k$) change during failure, which may change the region where the off-fault failure criterion is satisfied \citep{poliakov02, rice05}. In the simulations, the stress ratio $k$ was kept constant whereas it actually changed from $k = 2.2$ to $k = 1.8$ during quasi-static rupture. 3) The trajectory of the propagating rupture is not linear so that the principal stresses with respect to the trajectory of the rupture process zone change locally. 4) The applied stresses change orientation due to already formed fracture damage in and behind the rupture process zone \citep{faulkner06}. 

Some of the reasons above may be alleviated with a different loading geometry for better control on the principal stress orientations, using for instance a direct shear setup. More advanced rupture simulations with a `rough' rupture trajectory may yield additional insights into the lack of damage asymmetry. Nonetheless, the models suggest that the damage zone width for quasi-static and dynamic rupture in our experiments is controlled primarily by the rupture tip process zone. This is supported by the 3D $P$-wave velocity structure of the mixed rupture, where a low velocity zone is associated with the incipient secondary fault (Figure \ref{fig:6}). This added structural complexity provides a unique opportunity to compare damage in a fault zone without slip, with that in a fault zone that has accumulated 2.4~mm of slip (the main fault in the same sample). The lowest $V_\mathrm{P}$ in the zone around the incipient fault is only slightly higher than the lowest $V_\mathrm{P}$ around the fully developed fault (Figure \ref{fig:6}d), suggesting that rupture rather than slip caused the most off-fault damage. 

\subsubsection{Estimates for $\Gamma_\mathrm{off}$}
\begin{figure}
\centering
\includegraphics[scale = 0.9]{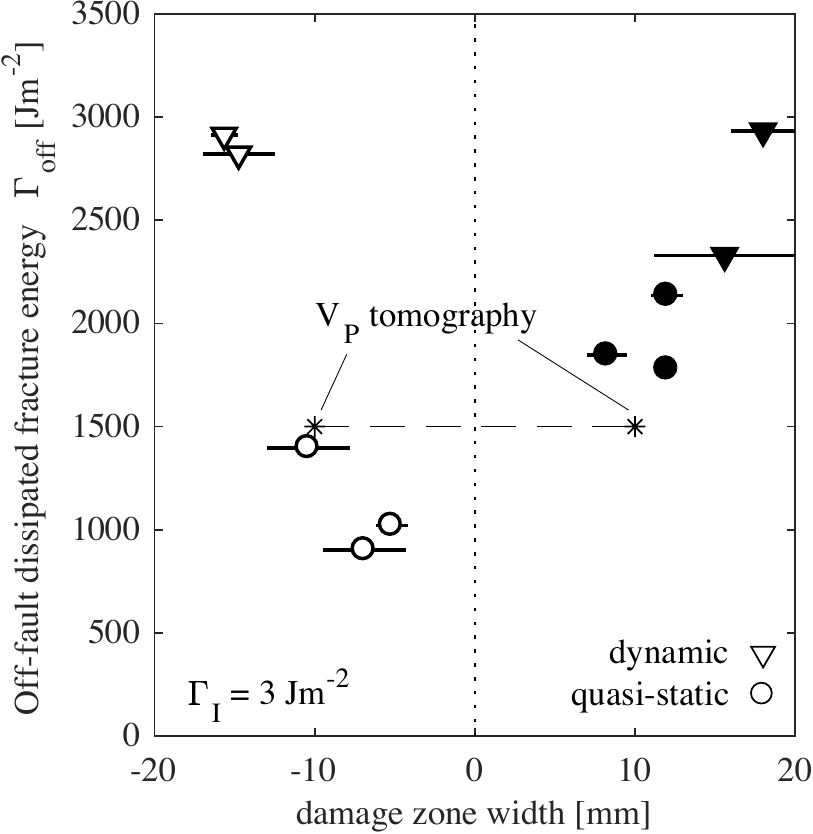}
\caption{$\Gamma_\textrm{off}$ (calculated using a specific surface energy of 3~Jm$^{-2}$) versus damage zone width. White symbols are transects on the tensile side of the fault, black symbols on the compressional side of the fault. The asterisks show the damage zone width and $\Gamma_\textrm{off}$ during quasi-static rupture, obtained from $P$-wave tomography by \citep{aben19}. }
\label{fig:GammaOff}
\end{figure}

A measure for $\Gamma_\mathrm{off}$ has been obtained in a previous study from \textit{in situ} $V_\textrm{P}$ tomography measurements on quasi-statically ruptured sample LN5 \citep{aben19}. We now use a second and independent method to obtain $\Gamma_\mathrm{off}$ from microstructural data for the same sample, and for a dynamically ruptured sample. The energy dissipated by creating new fracture surface in the volume around the fault gives an estimate for the off-fault dissipated fracture energy $\Gamma_\textrm{off}$. We assume that the microfractures are mostly tensile -- little to no slip and some opening of the microfractures observed in the thin sections testify to this -- so that the energy needed to form them is the mode I fracture energy. The cumulative mode I fracture energy gives us $\Gamma_\textrm{off}$, which was calculated from the fracture density data as follows:
\begin{linenomath*}
\begin{equation}
\Gamma_\textrm{off} = 2 \sum^{n}_{i = 1} \left( \rho^\textrm{frac}_{i} - \rho^{\textrm{frac}}_{0} \right) x_{i} \Gamma_\textrm{I}, 
\end{equation}
\end{linenomath*}
where $x_{i}$ is the fault-perpendicular width of image $i$ and $\Gamma_\textrm{I}$ is the mode I fracture energy for quartz and feldspar, ranging from 2 to 10 Jm$^{-2}$ \citep{atkinson87c}. A factor 2 is included to account for the two new surfaces that comprise each fracture. We calculated $\Gamma_\textrm{off}$ for all the fracture density transects obtained on both the quasi-statically ruptured and dynamically ruptured samples, for $\Gamma_\textrm{I} = 2$ and $\Gamma_\textrm{I} = 10$ Jm$^{-2}$. $\Gamma_\textrm{off}$ averaged for all transects for a quasi-static rupture ranges between 2 and 10~kJm$^{-2}$, that for a dynamic rupture is between 3 and 15~kJm$^{-2}$ (Figure \ref{fig:GammaOff}). $\Gamma_\textrm{off}$ for dynamic failure may be higher, considering that the fault damage zone width in the dynamically failed samples may be a lower bound only. $\Gamma_\textrm{off}$ increases nearly linearly with damage zone width, based on the transects through the tensile side of the fault (Figure \ref{fig:GammaOff}). For the quasi-static case, $\Gamma_\textrm{off}$ on the tensile side of the fault is less than on the compressional side of the fault.

The range of values for $\Gamma_\textrm{off}$ established by the microstructural approach depend mainly on the mode I fracture energy, for which we use values that vary by nearly an order of magnitude (2 and 10~kJm$^{-2}$), but are usually expected to be at the lower end of these value. We see a good agreement between the results from two independent methods to determine $\Gamma_\textrm{off}$: $\Gamma_\textrm{off}$ from $P$-wave tomography falls well within the estimated  range for $\Gamma_\textrm{off}$ from microstructures, and is similar to it when the mode I fracture energy is 3~Jm$^{-2}$ (Figure \ref{fig:11}c). 

We note that not all of the fracture surface area may have been traced from the SEM images, as we elected not to trace fractures below 9~$\mu$m since this would have increased noise (for instance, intrinsic flaws in the grains and artefacts from thin section preparation) more than increased actual fracture surface area. A close inspection of the SEM images (Figure \ref{fig:8}b-e) reveals that individual fractures shorter than 9~$\mu$m nearly all reside in zones of cataclasite and gouge close to the main failure zone, which have been excluded from further analysis. The microfractures further away from the fault zone are generally longer than 30~$\mu$m, and so the proportion of short microfractures not included as fracture surface area is small. This is confirmed by the good agreement between the two independent measures for $\Gamma_\textrm{off}$ at a realistic value for the mode I fracture energy. 

$\Gamma_\textrm{off}$ measured after dynamic failure is about 1.5 to two times higher compared to $\Gamma_\textrm{off}$ for quasi-static failure. This higher value for $\Gamma_\textrm{off}$ results from a wider damage zone and a higher overall fracture density. The first order estimate for damage zone width in section 4.3.2 suggests that the damage zone width is controlled by rupture velocity. A similar quantitative estimate for the cause of the difference in damage intensity cannot be achieved so easily. First, elastodynamic rupture models predict that with increasing rupture velocity, the state of stress in the rupture tip process zone exceeds the strength of the damage zone rock by an increasing amount \citep{poliakov02, rice05}, which may result in the formation of more microfractures. An increasing rupture velocity also increases off-fault strain rates that, when sufficiently high, give rise to a higher microfracture density due to inertia effects \citep{glenn86, liu98, bhat12, aben_AGUbook}. Second, off-fault stresses arising from slip along a rough fault will cause additional microfracturing and slip along off-fault microfractures formed during rupture. This effect could be represented in the rough fault model by decreasing the off-fault coefficient of friction, which also increases the distance at which slip along a rough fault interacts with rupture-induced off-fault damage. The microstructures of the dynamically failed sample indeed show a few small patches of fine material along secondary fractures at over 10~mm distance from the main fault, indicating that some slip occurred along this secondary fault. Energy dissipation by slip along off-fault microfractures is not considered in the above calculation of $\Gamma_\textrm{off}$.

\citet{moore95} observed peak fracture densities of the order of 40--80~mm/mm$^{2}$ in the microstructures of a quasi-static rupture propagation experiment on intact Westerly granite at 50~MPa confining pressure. Fracture densities dropped to a background density of around 14~mm/mm$^{2}$ at the damage zone boundary defined at 40~mm from the fault. Continuous microstructural observations were limited to 10~mm from the failure zone, except for one measurement at 40~mm distance. \citet{moore95} report values for $\Gamma_\textrm{off}$ that range from 1.7 to 8.6 kJm$^{2}$, which is similar to the values reported here (2 to 10 kJm$^{2}$, Figure \ref{fig:11}c). However, our results show a damage zone width after quasi-static failure of around 10~mm. The similarity in $\Gamma_\textrm{off}$ and the difference in damage zone width is not caused by a difference in resolution; both this study as well as \citet{moore95} have a cut-off for fractures smaller than 3~$\mu$m. Possible explanations for the difference in damage zone widths are: 1) Field and laboratory studies describe the evolution of fracture density in the fault damage zone by an exponential decay \citep{mitchell09, faulkner11b, moore95}, a logarithmic decay \citep{zang00}, or a powerlaw decay \citep{savage11, mayolle19, ostermeijer20} with increasing fault-perpendicular distance. This may result in different damage zone widths, but does not affect $\Gamma_\textrm{off}$ much as the 'tail' of the damage zone does not contribute significant amounts of additional fracture damage. 2) This study and \citet{moore95} used different granitic samples. 3) The confining pressure used by \citet{moore95} is half of that used in this study. Earthquake rupture simulations show that the damage zone width decreases with increasing confining pressure, while the relative damage intensity within the damage zone increases \citep{okubo19}. The results from this study and \citet{moore95} comply with these findings: An increase in confining pressure reduces the damage zone width while $\Gamma_\textrm{off}$ remains the same, which equals a higher microfracture intensity in a narrower damage zone.

\subsubsection{Off-fault dissipated energy and rupture energetics}
Is $\Gamma_\textrm{off}$ a significant energy sink for all preexisting fault in the brittle crust? The primary prerequisite for dissipation of fracture energy in the off-fault volume is that the imposed far-field stresses plus extraneous transient stresses in the rupture tip process zone are sufficiently high to damage the host material. In the experiments presented here, the failure zone material consists of the same material (intact granite) as the host rock, so that strength of the fault interface is the same as the strength of the surrounding material. The imposed stress state during rupture is thus high relative to the strength of the host rock. The magnitude of the additional transient stress field of the rupture tip process zone is proportional to $\Delta \sigma_{ij} \propto \Gamma^{1/2}$ for the limiting case of a singular shear crack \citep{freund90}, where $\Gamma$ is relatively high for intact granite. From these two arguments it follows that stresses around the experimental ruptures are high enough to induce off-fault damage, but should be considered an upper bound for pre-existing fault zones in terms of $\Gamma$ and strength. We can establish a lower bound scenario for a strong host rock and a weak interface, comprised off two bare granite slabs pressed together. Values of $\Gamma = 0.01-3.5$~Jm$^{-2}$ have been published for such an experimental setup \citep{ke18, kammer19}. These values are 5 to 7 orders of magnitude lower than for intact granite and were measured at 6~MPa normal stress, two orders of magnitude lower than our experiment. Thus both imposed far-field stress and the transient stress field are much lower than in our experiment, whereas the off-fault host material remains the same. We therefore expect no off-fault damage and a negligible value for $\Gamma_\textrm{off}$ in these experiments. These two cases mark the extremes for pre-existing faults, were our experiments are more illustrative for faults below 3~km depth where fault core materials likely experience rapid recovery of cohesion by sealing and healing processes, so that the fracture energy $\Gamma$ of the material increases sufficiently to entice damage in the host rock.

Fracture energy $\Gamma$ is a material parameter that is independent of fault slip and increases slightly with rupture velocity in most materials (i.e., the change in $\Gamma$ remains within the same order of magnitude) \citep{green74,freund90}. $\Gamma$ determined from mode I rupture experiments performed in PMMA and glass provide analogue results for mode II shear rupture experiments performed here. During mode I rupture in PMMA and glass, $\Gamma$ remains more or less constant below a critical velocity that is 0.36 (PMMA) or 0.42 (glass) of the Rayleigh wave speed, but increases by up to a factor 10 at higher rupture velocities up to the Rayleigh wave speed \citep{sharon96}. This increase in $\Gamma$ is an apparent one caused by microbranching instabilities along the main crack that creates additional fracture surface and accounts for the increase in $\Gamma$ \citep{sharon96}. At these rupture velocities, the single cracks still obey the initial $\Gamma$ measured at low rupture velocity \citep{sharon99}. Here, we show that part of the dependence of $\Gamma$ on rupture velocity is caused by an increasing amount of off-fault dissipated energy $\Gamma_\textrm{off}$. $\Gamma_\textrm{off}$ itself increases because the off-fault area in which energy is dissipated by microfracturing increases, and the amount of fractures within this area increases as well. What we measure as $\Gamma_\textrm{off}$ in our experiment is qualitatively similar to the additional energy dissipated by microbranching instabilities measured in PMMA during mode I rupture -- with the main difference that microbranching around a shear rupture in granite occurs already at quasi-static conditions as evidenced by the off-fault microfractures after quasi-static rupture. 

Fracture energy on the main failure plane ($\Gamma - \Gamma_\textrm{off}$) is partly invested as surface energy to create gouge and cataclasites, and partly dissipated as heat. We assume that fracture energy spent on the main failure plane does not change with increasing rupture velocity. $\Gamma$ thus only increases with rupture velocity if $\Gamma_\textrm{off}$ increases. In our experiments, $\Gamma_\textrm{off}$ doubles from around 3~kJm$^{-2}$ for quasi-static rupture to at least 5.5~kJm$^{-2}$ for dynamic rupture, and so $\Gamma$ increases by 10\%. An increase in $\Gamma$ means that ruptures will consume more energy to propagate, and a propagating rupture in a material with a velocity-dependent fracture energy will have a decreasing acceleration rate with increasing rupture velocity \citep{freund90}. 

Although the rupture velocity for the dynamic failure experiment is unknown, we can make a prediction for the evolution of $\Gamma$ if we adopt the simple relation that $\Gamma_\textrm{off}$ increases linearly with rupture-induced damage zone width. We observe this in our experiments (Figure \ref{fig:11}). We then take the relation between rupture velocity and damage zone width (Figure \ref{fig:13}), so that we can predict $\Gamma_\textrm{off}$. $\Gamma_\textrm{off}$ for rupture velocities near the Rayleigh wave speed increases by up to a factor of 10-20 relative to $\Gamma_\textrm{off}$ at low rupture velocity. Near the Rayleigh wave speed, we then expect that $\Gamma = 54-80$~kJM$^{-2}$. The factor 10-20 increase in $\Gamma$ is similar to that measured for PMMA. The critical velocity for a strong increase in $\Gamma$ for shear failure in granite under confinement is concurrent with the strong increase in damage zone width, at 0.81 of the Rayleigh wave speed (0.75 $V_{\textrm{S}}$) whereas the critical branching speed for PMMA is 0.36 in mode I rupture. 

\begin{figure}
\centering
\includegraphics[scale = 0.8]{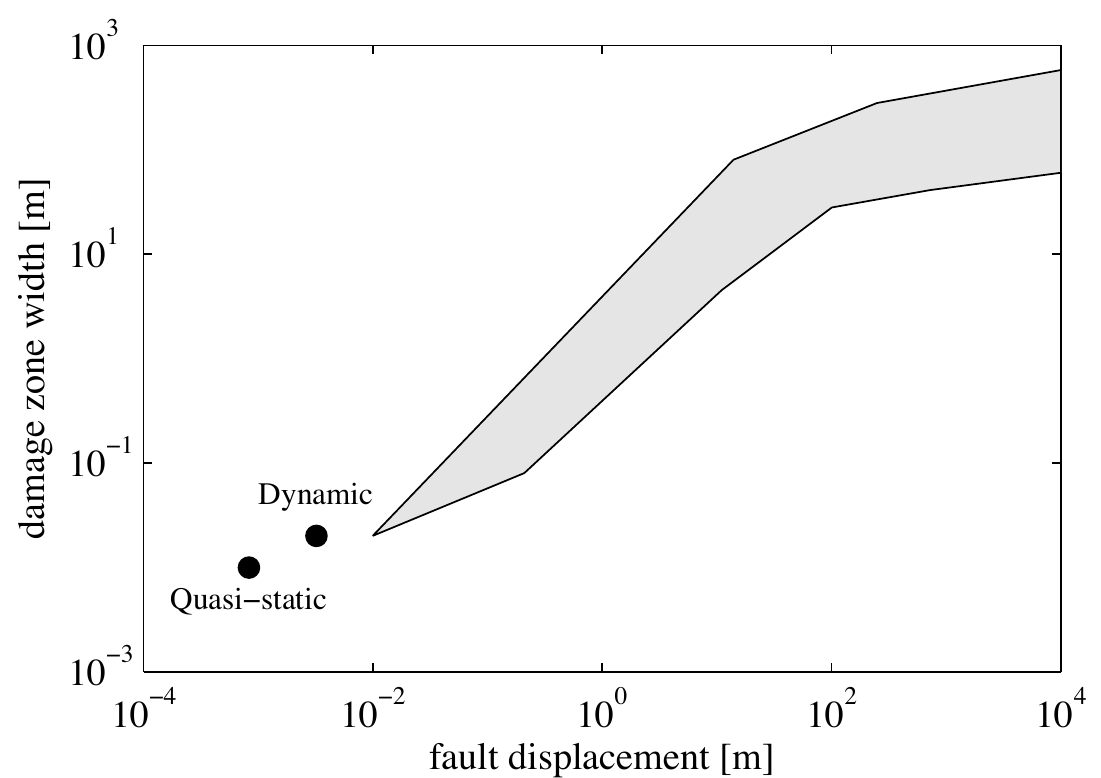}
\caption{Damage zone width versus total fault displacement, showing the damage zone width and slip from quasi-static rupture LN5 and dynamic rupture experiment LN7 (black datapoints). The shaded area shows the linear scaling relation between damage zone width and displacement, based on field data from \citet{savage11, faulkner11b}. }
\label{fig:14}
\end{figure}

Field observations show that damage zone width scales linearly with total fault displacement below 1.5-4~km \citep{shipton06b, savage11, faulkner11b} (Figure \ref{fig:14}). These studies argue that this relation is mainly due to slip-related off-fault damage by fault zone roughness and secondary faulting. By approximating off-fault stresses during rupture and during slip along rough faults, we show that for small displacements the rupture tip process zone determines the damage zone width (Figure \ref{fig:13}). Our observed damage zone widths after quasi-static (around 10~mm wide) and dynamic rupture (around 20~mm wide) confirm this: They are an order of magnitude larger than the slip that was accumulated during failure (0.83~mm slip for quasi-static rupture and around 3~mm slip for dynamic rupture), and thus do not fit with the linear scaling relation between damage zone width and slip (Figure \ref{fig:14}). 

 This is in contrast with what is argued by \citet{faulkner11b}, where it was suggested that the scaling relation goes through the origin (i.e., a zero displacement shear crack has no damage zone). Even at smaller negligible displacements, such as the failed secondary rupture in the mixed rupture experiment, a damage zone width is visible in the $P$-wave velocity structure (Figure \ref{fig:6}). We propose that the lower bound for the scaling relation observed in the field is determined by the stress field around a propagating rupture tip. The absolute value of this lower bound depends on the material properties, far-field stresses, and most importantly the rupture velocity. 

\citet{aben19} argued that the ratio between breakdown work $W_{\mathrm{b}}$ and its off-fault dissipated energy component is proportional to $\delta^{1-\lambda}$. $\lambda \approx 2$ for small earthquake slip below 10~cm, and $\lambda < 1$ for larger slip \citep{viesca15}, so that this ratio initially decreases with earthquake slip, to then stabilise or slightly increase with earthquake slip. For quasi-static failure and small slip ($<1$~mm for our quasi-static rupture experiment), all the breakdown work is spend as fracture energy, and so $\Gamma_\textrm{off}/\Gamma = 0.1$ \citep{aben19}. However, the scaling proposed by \citet{aben19} is based on the assumption that damage zone width increases linearly with fault slip as seen in the field \citep{faulkner11b, savage11}, whereas our results suggest that at very small amounts of slip the damage zone width is determined by rupture velocity (Figure \ref{fig:14} and \ref{fig:13}). The scaling relation between breakdown work and total off-fault dissipated energy is thus only valid when the damage zone width is determined by slip, i.e., fault roughness. 

\section{Conclusions}
We performed dynamic, quasi-static, and mixed shear failure experiments on Lanh\'elin granite to quantify the off-fault damage in the rupture tip process zones. The in situ $P$-wave structure and evolution was revealed by laboratory-scale seismic tomography during and after quasi-static failure and after dynamic failure. In both quasi-static and dynamic cases a localised low velocity zone formed around the fault interface, where a maximum reduction in $P$-wave velocity of about 25\% was observed. The low velocity zone around a fault created by dynamic rupture has a similar drop in $P$-wave velocities. The low velocity zones are caused by off-fault microfractures within the host rock around the fault during rupture and slip.
Using an effective medium approach, we computed microfracture densities from the $P$-wave tomography across the quasi-static and dynamic failure zones. The resulting theoretical microfracture densities are in good agreement with microfracture densities measured from thin sections, indicating that the $P$-wave tomography reveals realistic near-fault changes in elastic properties. We propose that a similar exercise using high resolution geophysical measurements combined with microstructural measurements from the field can reveal the physical properties around larger fault zones. 
The damage zone width established from microstructural analysis corresponds to the width of the low $P$-wave velocity zones, and is around 1~cm wide in the quasi-statically failed sample and 2~cm in the dynamically failed sample. Comparison with a previous microstructural study on quasi-static failed samples suggests that the damage zone width is depth dependent. We argue that the damage zone width in our experiment is controlled by rupture velocity and not by the slip up to a few mm. We propose that at larger slip the damage zone width is determined by fault roughness. Hence, in our experiments the increase in off-fault dissipated energy is mostly caused by an increase in rupture velocity. 
The off-fault dissipated energy $\Gamma_\textrm{off}$ that we measure is therefore associated to the fracture energy $\Gamma$, and was calculated from microstructural observations. $\Gamma_\textrm{off}$ increases from around 3~kJm$^{-2}$ for quasi-static rupture to at least 5.5~kJm$^{-2}$ for dynamic rupture, and shows that shear fracture energy in crystalline material increases with increasing rupture velocity. 

\begin{acknowledgments}
This study was funded by the UK Natural Environmental Research Council, grants NE/K009656/1 to N.B. and NE/M004716/1 to T.M.M. and N.B., and the European Research Council under the European Union's Horizon 2020 research and innovation programme (project RockDEaF, grant agreement \#804685 to N.B.). We thank J. Davy for thin section preparation and assistance with the SEM. We thank D. Kammer and an anonymous reviewer for their insightful comments. All data needed to evaluate the conclusions in the paper can be found at the NGDC repository of the British Geological Survey (https://www.bgs.ac.uk/services/NGDC) in dataset ID128186 for ultrasonic and mechanical data of sample LN5, and dataset ID135445 for all other data and SEM images. 
\end{acknowledgments}


\end{article}

\end{document}